%% file: cdwBreakdown.tex
\documentclass[pra,superscriptaddress,twocolumn,floatfig,showpacs,floatfix]{revtex4-1}

\usepackage{graphicx,color}
\usepackage{amsmath,amssymb,bm}
\usepackage{braket}		

\usepackage{mathtools}

\usepackage[plainpages=false,pdfpagelabels,colorlinks=true,
linkcolor=red,urlcolor=blue,citecolor=blue,pdftitle={Charge density wave breakdown a in heterostructure with electron-phonon coupling },pdfauthor={},pdfdisplaydoctitle=true,pdfduplex=DuplexFlipLongEdge]{hyperref}

\definecolor{darkred}{rgb}{0.90,0,0}
\definecolor{darkgreen}{rgb}{0,0.60,.2}
\definecolor{darkblue}{rgb}{0,0,1}
\definecolor{grey}{cmyk}{0,0,0,0.25}
\definecolor{orange}{cmyk}{0,0.6,1,0}

\usepackage{physics}
\begin{document}
\title{
Charge density wave breakdown in a heterostructure with electron-phonon coupling
}

\author{David Jansen}
\affiliation{Institut für Theoretische Physik,  Georg-August-Universit\"at G\"ottingen, D-37077 G\"ottingen, Germany}

\author{Christian Jooss}
\affiliation{Institut für Materialphysik,  Georg-August-Universit\"at G\"ottingen, D-37077 G\"ottingen, Germany}

\author{Fabian Heidrich-Meisner}
\affiliation{Institut für Theoretische Physik,  Georg-August-Universit\"at G\"ottingen, D-37077 G\"ottingen, Germany}

\begin{abstract}
Understanding the influence of vibrational degrees of freedom on transport through a heterostructure poses considerable theoretical and numerical challenges. 
In this work, we use the density-matrix renormalization group (DMRG) method together with local basis optimization (LBO) to study the half-filled Holstein model in the presence of a linear potential, either isolated or coupled to tight-binding leads.  In both cases, we observe a decay of  charge-density-wave (CDW) states at a sufficiently strong potential strength. Local basis optimization selects the most important linear combinations of local oscillator states to span the local phonon space. These states are referred to as optimal modes. We show that many of these local optimal modes are needed to capture the dynamics of the decay, that the most significant optimal mode on the initially occupied sites remains well described by a coherent-state typical for small polarons, and that those on the initially empty sites deviate from the coherent-state form.
 Additionally,  we compute the current through the structure in the metallic regime as a function of voltage. For small voltages, we reproduce results for the Luttinger parameters. As the voltage is increased,  the effect of larger electron-phonon coupling strengths becomes prominent.  Further,  the most significant optimal mode remains almost unchanged when going from the ground state to the current-carrying state in the metallic regime. 
\end{abstract}

\maketitle

\section{Introduction}
\label{sec:intro}

There is an increased technological and scientific importance of devices that are so small that quantum effects play an important role, 
see e.g.,~Refs.~\cite{Majumdar_04,shakouri_06,Giazotto_06,Rodgers_08}. Thus,  a detailed 
understanding of charge transport and the non-equilibrium properties of such quantum structures is desirable.  In particular, the formation of electron-phonon bound states, polarons, can give rise to the emergence of new electronic phases with electric transport characteristics different from metals or band semiconductors~\cite{franchini_21}. 

For example, at sufficient doping, charge-density waves (CDW) or charge-ordered (CO) states can evolve. Under strong electric fields, such states can break down, as, e.g., in CDW wires \cite{chavez_cervantes_19} or in CO manganites~\cite{polli_07}. In-situ transport experiments with CO manganites in an electron microscope revealed a complex transient behaviour with movement of entire CO domains and subsequent melting into a metallic state \cite{Jooss_07}.  Polaron transport, binding and dissociation in electric fields also have an impact on the performance of polymer solar cells \cite{Szarko_14}. Notably, CO can enhance the lifetime of polaron excitations in junctions and thus enable hot-polaron type solar cells \cite{kressdorf_20}. Understanding these experiments makes a comprehensive study of polaron transport and melting of CDW/CO states highly desirable. An improved and fully quantum-mechanical modelling of vibrational degrees of freedom \cite{galperin_2007,Andergassen_2010,Cuevas_2010,dubi_11,thoss_18} can help interpret the behaviour of molecules, see, e.g., Refs.~\cite{park00,smit_02,cocker_16}, and charge transport and thermalization in heterostructures, see, e.g.,  Refs.~\cite{ifland_15,Ifland_2017,kressdorf_20}.  

In general, CDW and CO states can hardly be distinguished in their ground states, since they have the same order parameter. However, the mechanism of forming the two ordered states is very different, i.e., a Peierls type lattice instability (CDW)~\cite{peirls_55} versus crystalization of localized charge carriers (CO)~\cite{Volja_2010,sotoudeh_17}. The difference is visible by the different phase transitions and non-equilibrium behaviour:  one the one hand, CDWs display metallic behaviour above their transition temperature. Below this temperature, collective phenomena~\cite{kida_02,wahl_03,cox_08,barone_09}, such as sliding, lead to non-linear behaviour in the current-voltage relation if a sufficiently strong electric field is applied. In CO systems, on the other hand, charge transport above the transition temperature is due to the hopping of localized polarons~\cite{Jooss_07,schmidt_08}.  The CDW states will be the main focus of this work and one of our goals is to better understand the CDW behaviour under an applied voltage.

From a theoretical point of view, one possible setup consists of a one-dimensional quantum structure sandwiched between two metals (see Fig.~\ref{fig:pic}).  Then, the transport properties of this structure can be investigated by applying a voltage difference to the two conducting leads.  Whereas the simplest case is to model a quantum dot with a certain energy level, many interesting and complicated extensions exist~\cite{Andergassen_2010,branschaedel_10,eckel_2010}.

One important example is to allow for local vibrational degrees of freedom on the quantum dot (see, e.g., Refs. ~\cite{galperin_04,galperin_2005,galperin_2006,galperin_2007,galperin_nitzan_07,Cuevas_2010,dubi_11,thoss_18}).  If these are modelled by a harmonic oscillator, one gets the single-level spinless Anderson-Holstein model (SAHM) which has been studied extensively in, e.g., Refs.~\cite{Koch_2010,jovchev_13,eidelstein_13, kherdi_17,kherdi_meden_17,khedri_18,shi_20,caltapanides2021}. One of the main goals of this work is to go beyond the SAHM and study a Holstein model, extending over several sites and coupled to leads.

Our main motivation is to get a better understanding of the effect of phonons on charge transport.  The one-dimensional Holstein model has a complex ground-state phase diagram (see, e.g., Refs.~\cite{bursill_98,creffield_05}) and, at half filling,  undergoes a transition  from  a Tomanga-Luttinger liquid (TLL) to a CDW  depending on the parameters. We address two main questions: How does the CDW phase behave when subjected to a linear potential and how do the different phases behave when the Holstein model is coupled to leads  and a bias voltage is switched on? 

The immense complexity of these types of non-equilibrium problems has driven the development of analytical and numerical methods such as the density-matrix renormalization group (DMRG) (see, e.g., Refs.~\cite{schmitteckert_04,al-Hassanieh2006,kirino_08, weichselbaum_09,Guo2009,branschaedel_10,HM_10,einhellinger_12,buesser_13,ganahl_2014,
dorda_15,schwarz_18,rams_20}), the numerical-renormalization group (NRG) (see, e.g., Refs.~\cite{wilson_75,krishna_80,anders_05,bulla_08}),  real-time renormalization group (RTRG)~\cite{schoeller_00},  functional-renormalization group (FRG)~\cite{schmidt_10,metzner_12,kherdi_17,kherdi_meden_17,khedri_18,caltapanides2021},  and quantum Monte Carlo (QMC) (as done in, e.g., Refs.~\cite{Han_07,werner_09,werner_10}).  In this work, we utilize DMRG for both ground-state search and time evolution~\cite{white92,daley2004,white04,vidal2004,schollwock2005density,schollwock2011density,paeckel_2019}. A comparison between several of these methods for the single-impurity Anderson model is contained in Ref.~\cite{eckel_2010}.

Here, we are particularly interested in modelling the influence of phonons,   requiring large local Hilbert spaces to capture the relevant physics in different parameter regimes.  Although this can, in principle,   be problematic for DMRG-based methods, techniques have been developed to treat these cases more efficiently (see also \cite{kloss_19,wall_16}).  One approach, introduced in Ref.~\cite{jeckelmann98}, consists of mapping the bosonic degrees of freedom onto pseudosites in the lattice,  thus replacing the large local Hilbert space with long-range interactions.  Another method consists of finding a basis where the local Hilbert space can be truncated significantly with negligible error~\cite{zhang98}, called the local basis optimization (LBO). Recently,  in Ref.~\cite{koehler20}, K{\"{o}}hler \textit{et al}.  suggested introducing bath sites,  thus treating a doubled system but with a restored U(1) symmetry.  The methodology  was further applied to time-evolution calculations in Ref.~\cite{mardazad_21}. These methods were benchmarked against each other for the ground-state search in Ref.~\cite{stolppkoehler20}.
For the problems at hand, we use LBO, which has already been successfully applied to a wide range of problems, e.g., in Refs.~\cite{zhang99,guo2012,brockt_dorfner_15,shroeder_16,brockt_17,stolpp2020,jansen20}.  

We first investigate the regular Holstein model (not coupled to leads) in the CDW phase, and focus on the breakdown of the charge order.  When we apply a linear potential,  the CDW order parameters defined in the electron and phonon sector decay rapidly if the inter-site potential difference is of similar magnitude as the polaron binding energy.  We further illustrate that the picture is similar to that in the Holstein dimer for short times.  Related studies have been done of the breakdown of a Mott insulator coupled to two leads in Ref.~\cite{HM_10},  of a Mott insulator due to an electric field in Refs.~\cite{oka_03,oka_05, oka_10,eckstein_10}, and of the Falicov-Kimball model, e.g., in Refs.~\cite{freericks_06,turkowski_07,freericks_08}.  Our work complements previous studies on the Holstein model where the CDW breakdown has been investigated by quenching parameters (e.g., in Ref. ~\cite{stolpp2020}) and applying a light pulse (e.g., in Ref.~\cite{hashimoto_ishihara_17}).  Additionally,  there have  been studies of single and bipolarons  in a linear potential (see,  e.g.,  Refs.~\cite{golez12a,vidmar_12}), and recently, on the heating of a CDW~\cite{weber_2021} and a CDW with pulsed electric fields~\cite{weber_Freericks_2021} and classical phonons.   

We then couple the Holstein model to leads.  Using time-dependent DMRG with LBO, we compute the current-voltage diagram of the structure in the metallic phase and demonstrate that using LBO leads to  a significant computational speed up. Note that the linear conductance of the model in the metallic phase has been computed in a similar set up in Ref.~\cite{bischoff_19}. There, the authors  used a Kubo-formalism based DMRG approach~\cite{Bohr_2006,bischoff_17}.  They also computed the Luttinger parameters ~\cite{apel_82,kane_92,kane_fischer_92} for the model and obtained quantitative agreement with Ref.~\cite{ejima09}, which computed the parameters from the structure factor.  We reproduce their results by analyzing the steady-state current at low voltages.    Further, we illustrate that the first optimal-basis state remains approximately constant in the current-carrying state as a function of time. 

In the CDW phase, large voltages are required for a clear decay of the order parameter to be seen at the time scales reached with our method.  Further,  the properties of the most significant optimal-basis state change in the initially empty site but remain well described by a coherent state in the initially occupied sites.

The main results of this paper can be summarized as follows:
In the charge-density wave regime of  the Holstein model, we simulate the CDW breakdown at large voltages. We explicitly demonstrate the decay of order parameters in both the phonon and electron sector. This is done for the Holstein model, as well as the Holstein model coupled to leads. Further, we show that the local phonon states of the initially empty sites in the CDW deviate from the coherent state, usually used to describe the small polaron.  In the metallic regime, we compute the current-voltage diagram and observe a significant dependence on the electron-phonon coupling at small voltages. We additionally illustrate that the local phonon distribution remains largely unaffected by  the voltage in the current-carrying state compared to the  ground  state and we reproduce literature values~\cite{ejima09,bischoff_17} for the Luttinger parameter. 

This paper is organized as follows: In Sec.~\ref{sec:model}, we introduce the model.  In Sec.~\ref{sec:meth}, we briefly review DMRG with LBO and how it can be applied to the systems studied here. In Sec.~\ref{sec:holgs}, we discuss CDW order in the ground state of the Holstein model with and without a coupling to the leads.  Section~\ref{sec:holmod} looks at the usual Holstein model and  Sec.~\ref{sec:vibstruct} at the Holstein model coupled to the leads. We summarize and give a brief outlook in Sec.~\ref{sec:conclusion}. 

\section{Model}
\label{sec:model}
\begin{figure}[t]            
    \centering              
    \def\svgwidth{200pt}    

    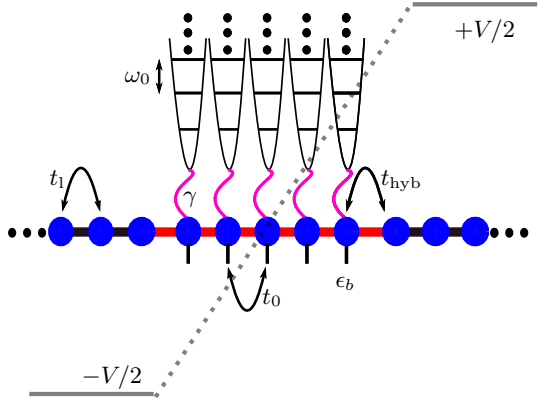  
    \caption{Two leads connected to a Holstein-model structure. The leads have hopping amplitude $t_{\textrm{l}}$.  In the structure, the electrons couple to the phonons with a coupling strength $\gamma$ and have a hopping amplitude $t_0$. The  phonons have the frequency $\omega_0$ and there is a local gate voltage $\epsilon_b$.  The electrons can tunnel between the leads and the structure with an tunneling amplitude $t_{\rm hyb}$.  At time $t=0$,  a voltage $V$ is applied, with a linear interpolation through the structure.}  
    \label{fig:pic}        
\end{figure}
    
When investigating how vibrating degrees of freedom affect the transport properties of a structure attached to two leads we  can write down a three-term model:
\begin{equation} \label{eq:def_Hhet}
\hat H = \hat H_{\rm leads}+ \hat H_{\rm hyb}+ \hat H_{\rm s}.
\end{equation}
We use open-boundary conditions and set $\hbar=1$ throughout this paper.
If the structure starts at site $L_0$ and is of length $L_{\rm s}$,  the Hamiltonian of the leads becomes
\begin{equation}
  \label{def_Hlead}
\begin{split}
  \hat H_{\rm leads} & = -t_{\rm l} \sum\limits_{j=1}^{L_0-2} \left( \hat c_j^{\dag} \hat c_{j+1}^{\phantom{\dag}} +  \hat c_{j+1}^{\dag} \hat c_j^{\phantom{\dag}}        \right)  \\
& -t_{\rm l} \sum\limits_{j=L_0+L_{\rm s}}^{L-1} \left( \hat c_j^{\dag} \hat c_{j+1}^{\phantom{\dag}} +  \hat c_{j+1}^{\dag} \hat c_j^{\phantom{\dag}}        \right),
\end{split}
\end{equation}
where  $\hat c^{(\dag)}_j $ is the electron annihilation (creation) operator on the $j$-th  site, $L$ is the total length of the system, and $t_{l}$ the hopping amplitude in the leads.
The hybridization term is
\begin{equation} \label{def_Hhyb_dot}
\begin{multlined}
\hat H_{\rm hyb}=-t_{\rm hyb} \left( \hat c_{L_0-1}^{\dag} \hat c_{L_0}^{\phantom{\dag}} +  \textrm{H.c.}     \right) \\
-t_{\rm hyb} \left( \hat c_{L_0+L_{\rm s }-1}^{\dag} \hat c_{L_0+L_{\rm s}}^{\phantom{\dag}} +  \textrm{H.c.}     \right),
\end{multlined}     
\end{equation}
and $t_{\rm hyb} $ is the hopping amplitude between the structure and the leads. 
To model vibrational degrees of freedom in the structure, we choose the Holstein model~\cite{Holstein1959}, which contains a coupling between electrons and local optical phonons. 
The structure's Hamiltonian then takes the form:
 \begin{multline} \label{def_Hstruct}
\hat H_{ \rm s} = - t_{0}  \sum\limits_{j=L_0} ^{L_0+L_{\rm s}-2}\left( \hat c_{j}^{\dag} \hat c_{j+1}^{\phantom{\dag}} + \textrm{ H.c.}     \right) 
\\  +\sum\limits_{j=L_0}^{L_0+L_{\rm s}-1} ( \omega_0 \hat b_j^{\dag} \hat b_j^{\phantom{\dag}} + \gamma \hat n_j (\hat b_j^{\dag} + \hat b_j^{\phantom{\dag}}) + \epsilon_b \hat n_j),
\end{multline}
with $\hat b^{(\dag)}_j $ being the phonon  annihilation (creation) operator on site $j$ and $\hat n_j=\hat c^{\dagger}_j \hat c_j$. Further, we have the gate voltage $\epsilon_b$, the harmonic-oscillator frequency $\omega_0$ and the coupling strength between the electrons and the phonons  $\gamma$.  For $L_{\rm s }=1$, the model turns into the well-studied spinless Anderson-Holstein model. The complete model is illustrated in Fig.~\ref{fig:pic}. 
We also define 
\begin{equation} \label{def_epstilde}
\tilde{\epsilon}=\epsilon_b - \frac{\gamma^2}{\omega_{0}},
\end{equation}
and in this work, we set $\tilde{\epsilon}=0$ so that $\epsilon_b$ corresponds to the polaron binding energy in the single-site limit of the model.  In this regime, we can detect a clear distinction between the metallic and CDW phase in the ground state for the parameters investigated here. When we refer to the regular Holstein model, we mean the Hamiltonian in Eq.~\eqref{def_Hstruct} without leads.

When studying the Holstein structure coupled to leads, we always start with the ground state of the Hamiltonian in Eq.~\eqref{eq:def_Hhet}, and at time $t\omega_0>0$, we apply a voltage by adding the term
\begin{equation} \label{def_structHV}
\begin{split}
\hat H_{\rm V} = \frac{-V \theta(t)}{2} \sum\limits_{j=1}^{L_0-1} \hat n_j +\frac{V \theta (t)}{2} \sum\limits_{j=L_0+L_{\rm s}}^{L} \hat n_j \\
 + \sum\limits_{j=L_0}^{L_0+L_{\rm s}-1} \theta (t)(i-L_x)\Delta V  \hat n_j
\end{split}
\end{equation}
to the Hamiltonian. Here, $\Delta V=V/(L_{\rm s}+1)$ and $L_x=L_0-1+(L_{\rm s}+1)/2$.
For the regular Holstein model, we only apply the linear potential [i.e., the last term in Eq.~\eqref{def_structHV}].

We further define the hybridization parameter
\begin{equation} \label{def_GamHV}
\Gamma= 2 (t_{\rm hyb })^2.
\end{equation}

We are interested in the expectation values of several observables. We calculate the expectation value of the current through the structure defined as 
\begin{equation} \label{def_curr}
\hat j=  \frac{i }{2} t_{\rm hyb}\left( \hat c_{L_0-1}^{\dag} \hat c_{L_0}^{\phantom{\dag}} -  \textrm{H.c.}    + \hat c_{L_0+L_{\rm s}-1}^{\dag} \hat c_{L_0+L_{\rm s}}^{\phantom{\dag}} -  \textrm{H.c.}     \right),
\end{equation}  
where we take the average of the incoming and outgoing currents.
Additionally, we compute an order parameter in the electron sector
\begin{equation} \label{def_Opamn}
 \mathcal{O}_{ n}=\frac{1}{N_e} \sum\limits_{i=L_0}^{L_0+L_{\rm s}-1} (-1)^{i-L_0} \expval*{\hat  n_i},
\end{equation}
and in the phonon sector
\begin{equation} \label{def_OpamX}
 \mathcal{O}_{X}=\frac{-1}{N_e} \sum\limits_{i=L_0}^{L_0+L_{\rm s}-1} (-1)^{i-L_0} \expval*{\hat X_i},
\end{equation}
 where $\hat X_i= \hat b^{\dagger}_i+\hat b_i$ and $N_e=(L_{\rm s}+1)/2$ for odd $L_{\rm s}$.  In Eq.~\eqref{def_OpamX}, we include an additional minus sign to ensure that  $\mathcal{O}_{ X}>0$ in the groundstate for $\gamma>0$.  These parameters characterize the transition from a TLL to a CDW phase. 

\section{Methods}
\label{sec:meth}
In this section, we briefly explain the main numerical method used in this work, namely the time-dependent density-matrix renormalization group using LBO. DMRG-based methods \cite{white92,schollwock2005density,schollwock2011density}  have  proven to be an extremely valuable tool to study one-dimensional systems and have already been applied extensively to a wide range of problems (see  Refs.~\cite{schollwock2011density,schollwock2005density,paeckel_2019} for reviews). This work uses time-dependent DMRG~\cite{daley2004,white04, vidal2004} with local basis optimization~\cite{zhang98}.  LBO has been combined with both exact-diagonalization methods \cite{zhang99} and matrix-product state methods~\cite{brockt_dorfner_15,shroeder_16,brockt_17,stolpp2020,jansen20} and has enabled the study of electron-phonon systems in previously inaccessible regimes for other wave-function based methods, e.g., for finite-temperature spectral functions of the Holstein polaron~\cite{jansen20},  quench dynamics of charge-density waves with a completely quantum mechanical treatment of the phonons~\cite{stolpp2020}, and the scattering of an electronic wave packet on a structure with electron-phonon interaction~\citep{brockt_17}.
Other Hilbert-space based methods used in the field are
exact diagonalization, see, e.g., in Refs.~\cite{zhang99,capone_97,jansen19}, diagonalization in a limited functional space, e.g., in Refs.~\cite{bonca99,golez12a,dorfner_vidmar_15}, and the Lanczos method, e.g., in Refs.~\cite{wellein96,wellein98,bonca2019}.

Here,  for the Holstein model coupled to leads, we use matrix-product states consisting of both sites with only fermionic (indicated by $\sigma$) and  sites with both fermionic and bosonic degrees of freedom (indicated by $\eta$). We write $\ket{\vec{\sigma}}_{\rm left}=\ket{\sigma_1, \hdots ,\sigma_{L_0-1} }$,  $\ket{\vec{\sigma}}_{\rm right}= \ket{\sigma_{L_0+L_{\rm s} } , \hdots , \sigma_{L} }$ and  $\ket{\vec{\eta}}_R= \ket{\eta_{L_0}, \hdots , \eta_{L_0+L_{s} -1}}$. We truncate the phonon Hilbert space by allowing maximum $M$ phonons on each site.  The total matrix-product state can be written as:
 \begin{multline} \label{eq:MPS1}
     \ket*{\psi}=\sum\limits_{ \ket{\vec{\sigma}}_{\rm left}, \ket{\vec{\sigma}}_{\rm right} , \ket{\vec{\eta}}} 
    A^{ \sigma_1}   \hdots  A^{\sigma_{L_0-1}}  A^{\eta_{L_0}} \hdots \\  \hdots  A^{\eta_{L_0+L_{\rm s}-1}}  A^{\sigma_{L_0+L_{\rm s }}}\hdots A^{ \sigma_{L}}
     \\
    \ket{\vec{\sigma}}_{\rm left} \ket{\vec{\eta}} \ket{\vec{\sigma}}_{\rm right}.
 \end{multline}
  
  For the time evolution,  the Hamiltonian is first written as a sum of terms  $\hat h_{l}$ acting on the two neighbouring sites $l$ and $l+1$. For a time step $dt$,  we  carry out  a second-order Trotter-Suzuki decomposition into even and odd terms
     \begin{equation} \label{eq:trot}
e^{-idt\hat H}=e^{-idt  \hat H_{\rm even} /2}e^{-idt \hat H_{\rm odd}}e^{-idt \hat H_{\rm even} /2}+ O(dt^3)\, .
\end{equation}
The gates can now be applied directly to the matrix-product state.

To treat the large number of local degrees of freedom efficiently we apply a transformation into a local optimal basis. This is done by obtaining the local reduced density matrix $\rho$ at each time step after applying the time-evolution gate to a site with bosonic degrees of freedom.  As described in detail in Refs.~\cite{brockt_dorfner_15,brockt_17,stolpp2020,jansen20}, this is used to obtain the transformation matrices into the new, called the optimal local basis by diagonalizing $\rho$ such that 
     \begin{equation} \label{eq:rho_diag}
\rho=U^{\dagger} W U.
\end{equation}
In Eq.~\eqref{eq:rho_diag}, $W$ is a diagonal matrix containing the eigenvalues $w_{\alpha}$ and $U$ is the transformation matrix with the eigenvectors $\ket{\phi_{\alpha}}$ such that  
     \begin{equation} \label{eq:eigveceq}
\rho \ket{\phi_\alpha}= w_{\alpha} \ket{\phi_\alpha}.
\end{equation} The matrices $U$ transform  between the phonon bare mode basis and the optimal basis on a given site. The number of optimal states one needs to keep while allowing for a certain error is set by the eigenvalues $w_{\alpha}$. In the single-site limit  of the Holstein model,  it is well known that the system can be described by only two optimal states,  the coherent and the empty state. In this case, the optimal basis has an obvious physical meaning.  Since the system conserves the number of electrons we can split up the reduced density matrices into block matrices in the one-electron sector $\rho^1$ and the zero-electron sector $\rho^0$ with the sum of the traces $\mbox{Tr}[\rho^1]+\mbox{Tr}[\rho^0]=1$. We further denote the corresponding eigenvalues and eigenvectors with an additional index so that, e.g., $ \rho^1 \ket*{\phi^1_\alpha}= w_{\alpha}^1 \ket*{\phi^1_\alpha}$. The weights $w_{\alpha}^1$ in our system are analysed in 
Appendix~\ref{sec:app1}.  One important result of this work is that the local basis optimization works very well for the calculations done in the metallic phase (only $\sim 3$ local states are needed in one particle number sector,  compared to $M+1$ (in this case $M=30$) in the bare phonon-number basis). There, only relatively small voltages are needed to obtain a steady state current and a current-voltage diagram. In contrast,  in all cases where the CDW is found to break down, large voltages and many local states are required. 

In the time-dependent DMRG method with LBO we first apply the time-evolution gate, then obtain the transformation matrices and transform into the optimal basis before the subsequent singular value decomposition. A thorough discussion of the method can be found in Ref.~\cite{brockt_dorfner_15}.

When diagonalizing the reduced density matrix to obtain the optimal basis,  the smallest eigenvalues $w_{\alpha}$ are discarded such that the truncation error is below a threshold: $\sum\limits_{  \rm{discarded} \; \alpha} w_{\alpha}/(\sum\limits_{\rm{all} \: \alpha} w_{\alpha})<\epsilon_{\rm LBO}$. For the truncation done in the time-evolution scheme, we discard all singular values such that $\sum\limits_{ \rm{discarded } \; \alpha} s_{\alpha}^2/(\sum\limits_{\rm{all} \:  \alpha } s_{\alpha}^2)<\epsilon_{\rm bond}$.  
All calculations were done using Ref.~\cite{itensor} and with $dt\omega_0=0.025$.
 
\section{ CDW order in the ground and initial state }
\label{sec:holgs}

\begin{figure}[t]
\includegraphics[width=0.99\columnwidth]{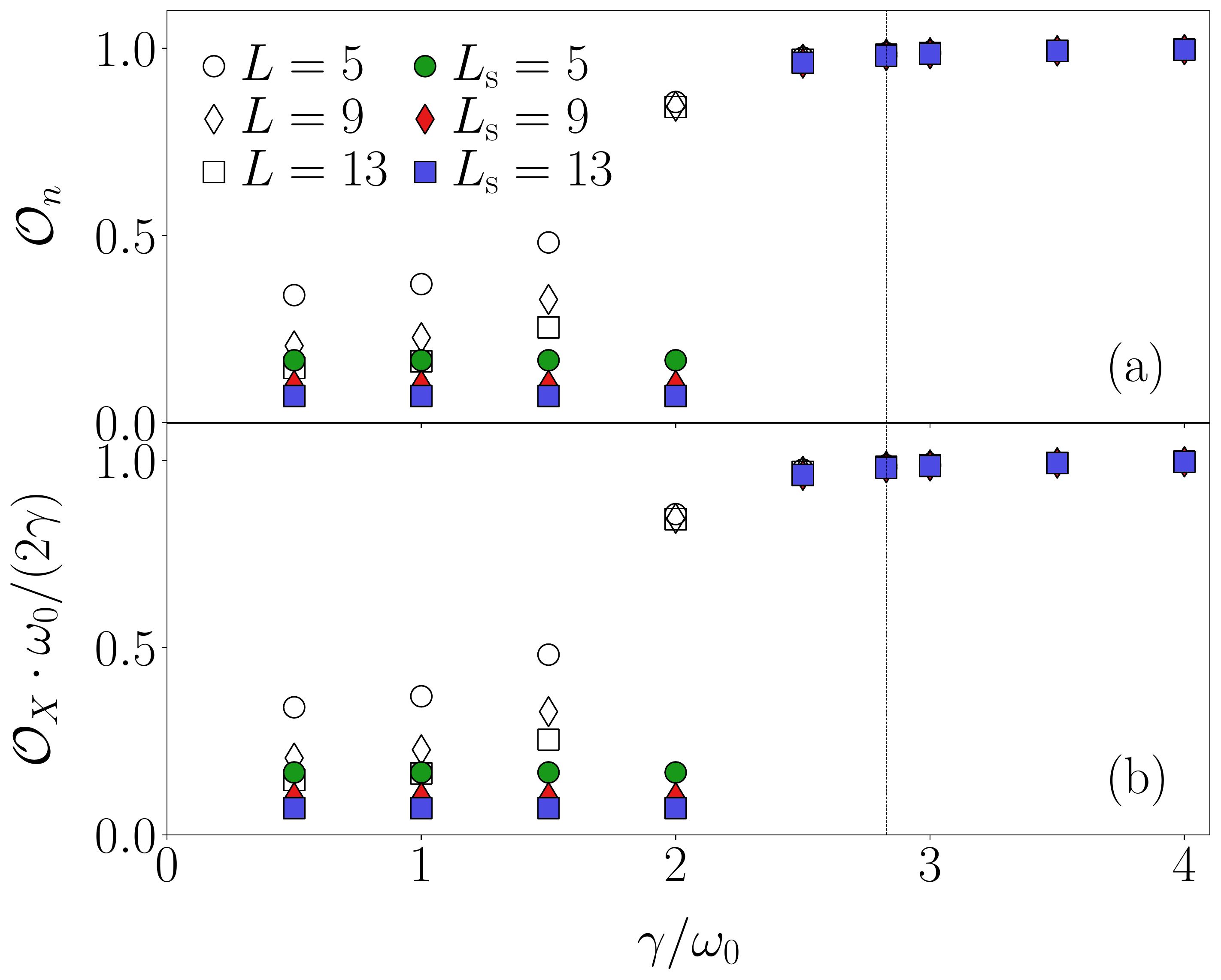}
\caption{(a) Order parameter in the electron sector for the ground states of the Holstein model (open symbols) and the Holstein model coupled to leads (filled symbols) for different $\gamma/\omega_0$.  $L$ refers to the system size of the Holstein model and $L_{\rm s}$ to the length of the structure which is coupled to the leads.  In both cases, we set $t_0 / \omega_0=1,\tilde{\epsilon}=0$, $M=35$ for $\gamma/\omega_0\leq2$, and  $M=50$ for $\gamma/\omega_0>2$.  For the Holstein model coupled to leads, we further use a total $L=236, t_{l}/\omega_0=2$ and $ \Gamma /\omega_0=1$.  The dashed line indicates $\gamma/\omega_0=2\sqrt{2}$, which is predominately used later in this work.  (b) Order parameter in the phonon sector for the same parameters as in (a).  Note that we do not show the data point for $L_{\rm s}=5$, and $\gamma/\omega_0=2.5$, since it is not converged conclusively with respect to the criteria in the main text. }
\label{fig:orderparam_diffmodels}
\end{figure}

We first look at the ground-state properties of the two setups.  In  Fig.~\ref{fig:orderparam_diffmodels}, we show $\mathcal{O}_{ X}$ and $\mathcal{O}_{ n}$ for the ground state obtained with DMRG for both the Holstein model and the Holstein model coupled to leads.  The relative variance of the ground-state energy, $\sigma_E^2=(\expval*{H^2}-\expval*{H}^2)/\expval*{H}^2$, is converged up to the order $\leq 10^{-6}(10^{-12})$ for the regular Holstein model (Holstein model coupled to leads).  We further verify that the obtained state is robust with respect to different initial states. The data are shown for both the structure coupled to leads and the regular Holstein model for different $\gamma/\omega_0$ with $M=35$ local phonon states for $\gamma / \omega_0\leq2$ and $M=50$ for  $\gamma / \omega_0>2$. One can observe a clear distinction between the charge-density-wave phase and the metallic phase.  For small $\gamma / \omega_0$,  the order parameters decrease as the system size is increased. Note that in the metallic case, the values for the Holstein model coupled to leads are smaller than for the Holstein chain itself at the same length since for the structure, the total system size is even while the number of sites in the structure is odd. As $\gamma / \omega_0$ is increased,  both $\mathcal{O}_{ n}$ and $\mathcal{O}_{ X}\cdot \gamma / \omega_0$  approach a constant for all systems, indicating the charge-density-wave phase. We further confirm that the results remain consistent as the maximum local phonon occupation number $M$ is increased. 

\section{Holstein model with a linear potential}
\label{sec:holmod}

\begin{figure}[t]
\includegraphics[width=0.99\columnwidth]{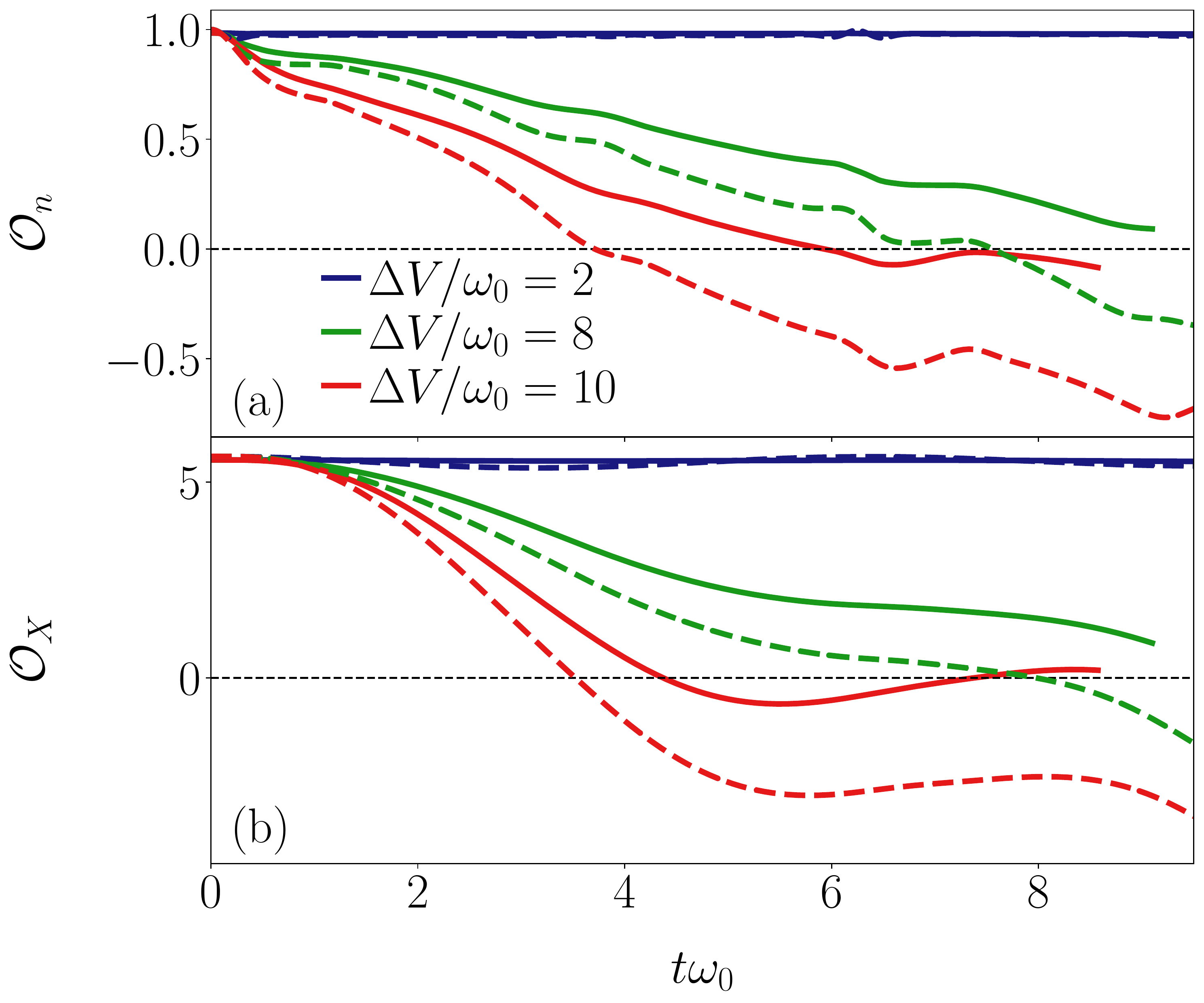}
\caption{Order parameters  for the Holstein model with $L=9, \gamma / \omega_0=2\sqrt{2},  \tilde{\epsilon}=0,t_0/\omega_0=1,M=50$ and different voltages $V / \omega_0$ ($\Delta V$ is the potential difference between consecutive sites in the structure).   For the calculations, we use $\epsilon_{\rm LBO}=10^{-7}$ and $\epsilon_{\rm bond}=10^{-7}$. (a) Order parameter in the fermion sector, see Eq.~\eqref{def_Opamn}.  (b) Order parameter in the bosonic sector, see Eq.~\eqref{def_OpamX}.  The dashed lines are the exact data for the Holstein dimer.  }
\label{fig:HolOP}
\end{figure}

In this section, we focus on the Holstein model without coupling to any leads [i.e., just Eq.~\eqref{def_Hstruct}]. 
To investigate the dynamics of the CDW, we initially compute the ground state of the Holstein model and at  time $t\omega_0=0$ we apply the linear potential. We focus on the strong-coupling regime, $\gamma /\omega_0=2\sqrt{2},t_0/\omega_0=1$, where the ground state is known to be a charge-density wave \cite{bursill_98,creffield_05} (see Fig.~\ref{fig:orderparam_diffmodels}).  In Fig.~\ref{fig:HolOP}, we show the order parameters for different voltages. Since a single polaron has a binding energy of $\epsilon_b= \gamma^2/\omega_0$, one expects that local voltage differences of $\Delta V / \omega_0 \sim \gamma^2/\omega_0^2$ are needed for the order parameter to decay on the time scales reachable here.  This is confirmed for both the order parameter in the electron [Fig.~\ref{fig:HolOP}(a)] and phonon sector [Fig.~\ref{fig:HolOP}(b)]. There, we observe almost no decay for $V / \omega_0=20$, which corresponds to $\Delta V/\omega_0 =2 \ll\gamma^2/\omega_0^2=8 $. However, for both $V / \omega_0=80 $ $(\Delta V/\omega_0=8)$ and $V / \omega_0=100$ $ (\Delta V/\omega_0=10)$, the order parameters decay substantially.  Since one could expect that only the voltage difference between two neighbouring sites should dominate the order-parameter decay, we also show the data for the same voltage differences in the Holstein dimer.  The dynamics are indeed similar, indicating that the dimer picture gives a reasonable qualitative description for the short-time dynamics.
  
\begin{figure}[t]
\includegraphics[width=0.99\columnwidth]{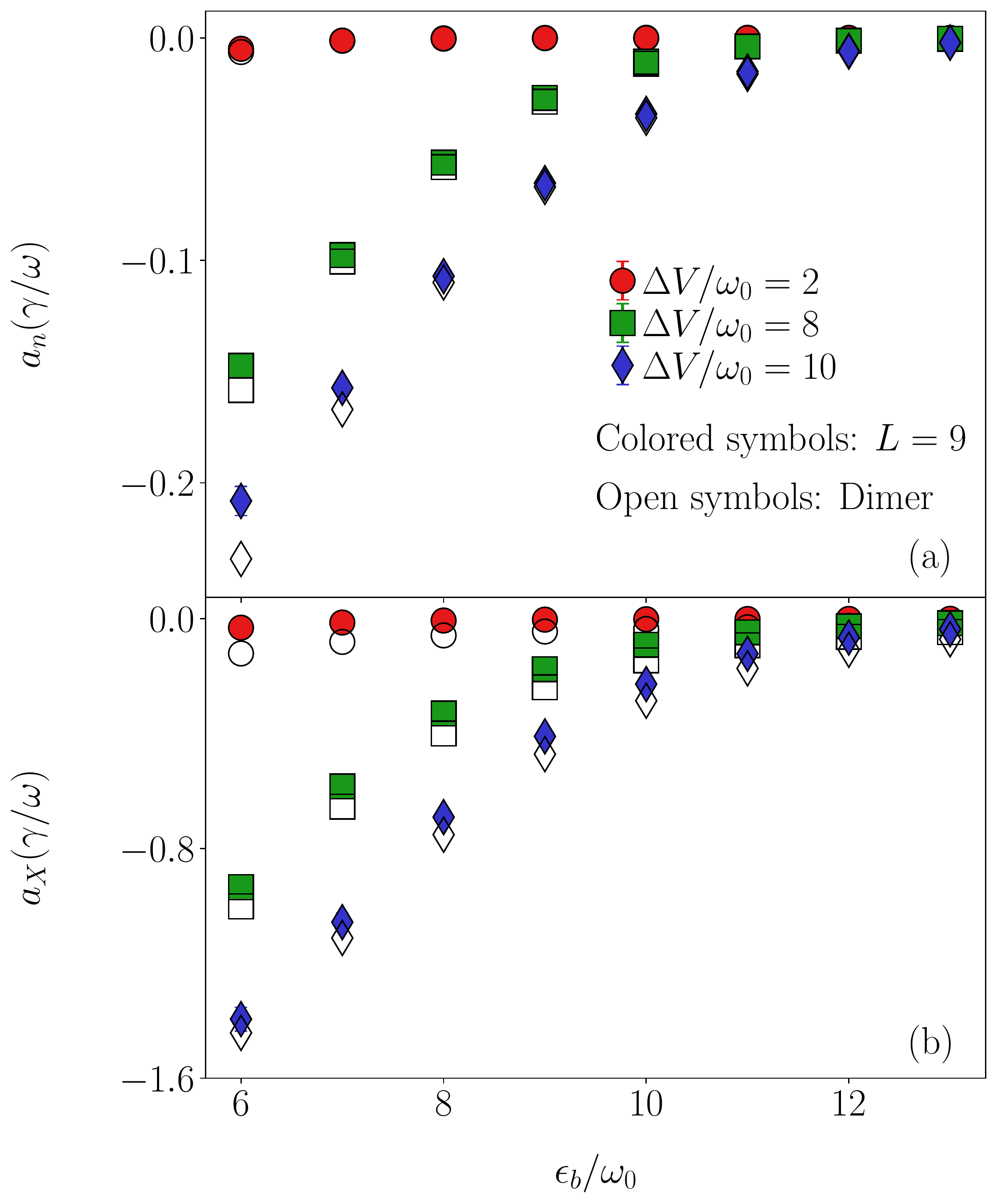}
\caption{Slopes of the linear fit to the initial decay of the order parameters (see main text for details) in the Holstein model. We use $L=9, \gamma / \omega_0=2\sqrt{2},  \tilde{\epsilon}=0,t_0/\omega_0=1,M=50$ and different voltages $V / \omega_0$. For the calculations, we use $\epsilon_{\rm LBO}=10^{-7},\epsilon_{\rm bond}=10^{-7}$.  The error bars indicate the standard deviation of the errors with maximum values of the order of $10^{-2}$ and thereby can not be seen for most points on the scale of the figure. Note that the $L=9$ data are rescaled with the factor 5/4 to make the number of electrons participating in the CDW breakdown commensurate with the Holstein dimer.}
\label{fig:coupdens}
\end{figure}

 To quantify how the change of the order parameter is affected by the electron-phonon coupling strength, we fit $\mathcal{O}_{ n}$ and $\mathcal{O}_{ X}$ in the interval $t \omega_0 \in [0, 6]$ with the function $f(t)=at+b$.
 In Fig.~\ref{fig:coupdens}, we show the resulting values of $a$ as a function of $\epsilon_b =\gamma^2/\omega_0$ for different voltages for the Holstein model and the dimer.  The figure indicates that the decrease of the order parameters strongly depends on the polaron binding energies and the inter-site voltage $\Delta V$.   In all cases, the decay gets suppressed when the binding energy gets large.  Further, the figure illustrates that the functional dependence of the $a$'s are similar for both the Holstein model and the dimer.    To conclude, our results suggest that a breakdown of the CDW can already be seen for $\Delta V\approx\epsilon_b$, but 
gets more prominent for  $\Delta V \gg \epsilon_b$. For $\Delta V \ll \epsilon_b$, the CDW is stable.

\section{ Holstein structure coupled to leads}
\label{sec:vibstruct}
In this section, we move on to study the Holstein structure coupled to leads with $L_{\rm s}=9$.  In the first part, we focus on couplings strengths which lie in the TLL regime before we go to the charge-density wave regime.
\subsection{Metallic phase}
\label{subsec:metal}
\subsubsection{Currents}
\label{subsubsec:currents}
We first compute the current-voltage curve for the model with coupling strengths $\gamma/\omega_0$ in the TLL regime.  To do this, we apply the commonly used technique, see, e.g., Refs.~\cite{kirino_08, branschaedel_10,HM_10}, of averaging the expectation values of $\expval*{\hat j(t) }$ in a time interval where a quasi-steady state current  is reached.  We call this quantity $\expval*{\hat j(t)} _{\rm av}$. We choose the interval $t\omega_0 \in [20,30]$.  Typical data for the time dependence of the current $\expval*{\hat j(t) }$ is shown in Appendix~\ref{sec:app1}.
For our data, the standard deviation is defined as \begin{equation}
\sigma_{\rm STD} (\expval*{\hat j})=\sqrt{\sum_{t_i\omega_0 \in [20,30] } \frac{1}{N} |\expval*{\hat j(t_i) }-\expval*{\hat j(t)} _{\rm av}|^2},
\end{equation} where $t_i$ is a point in time depending on the time step, $N$ is the number of terms in the sum $\sum_{t_i\omega_0 \in [20,30] }$, and we have $\sigma_{\rm STD} (\expval*{\hat j})/\expval*{\hat j(t)} _{\rm av}$of order $\leq 10^{-3}$.  

\begin{figure}[t]
\includegraphics[width=0.99\columnwidth]{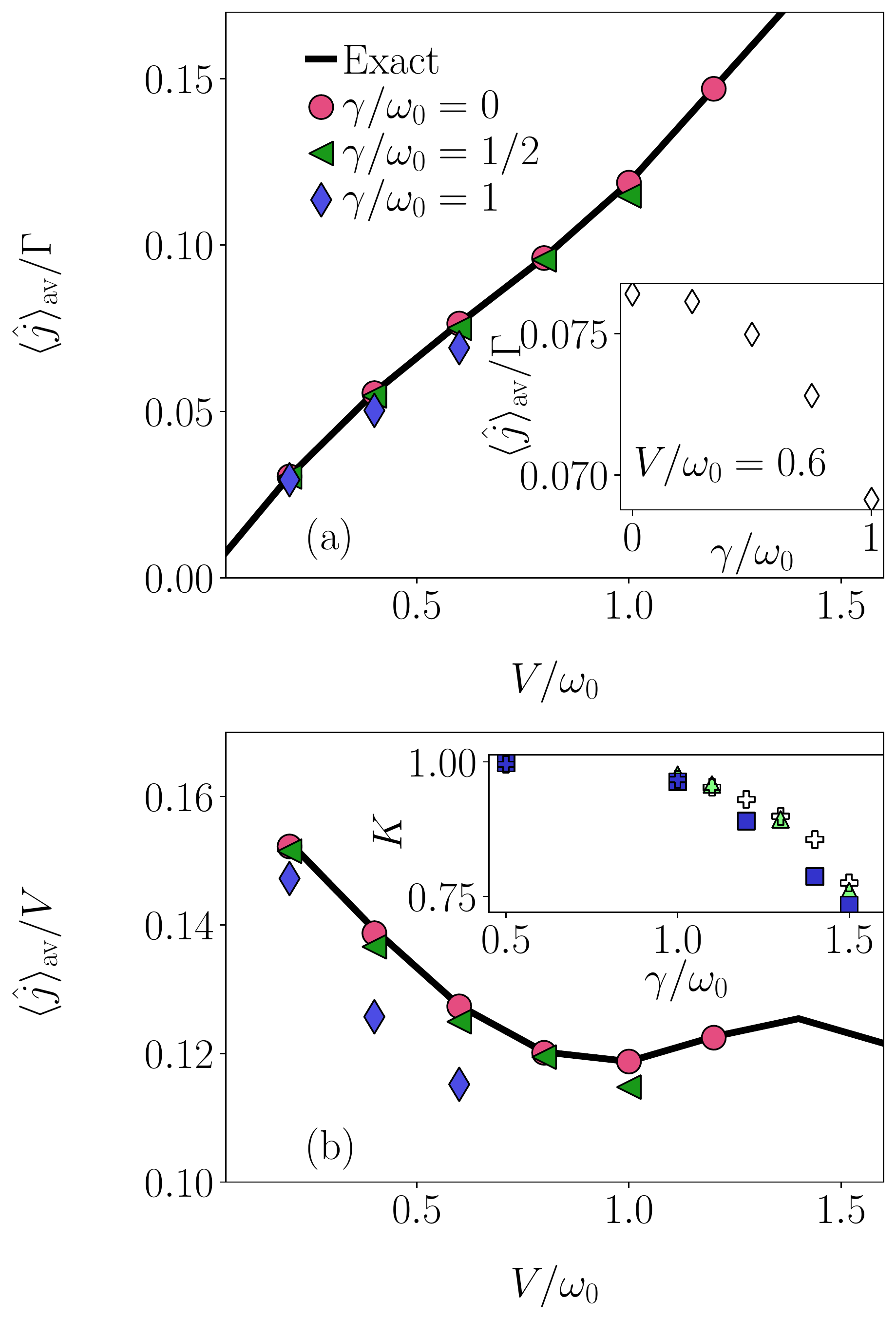}
\caption{(a) Average current $\expval*{\hat j}_{\rm av}$,  see text for details,  for the structure with $L_{\rm s}=9,L=236,  \Gamma /\omega_0=1, t_{l}/\omega_0=2,\epsilon_{\rm bond}=10^{-8}, \epsilon_{\rm LBO}=10^{-7}, \tilde{\epsilon}=0$ and different $\gamma / \omega_0$ in the metallic phase.  We further show exact results for $\gamma / \omega_0=0$ (black solid line).  The inset in (a) shows $\expval*{\hat j}_{\rm av}$ at fixed  $V/ \omega_0=0.6$ as a function of $\gamma / \omega_0$.   (b) Same data as in (a) but divided by $V$. The inset in (b) shows the Luttinger-liquid  parameter, see main text for details, together with values obtained with different methods. The plus signs are calculated from our data,  the green triangles are from Ref.~\cite{ejima09}, and the blue squares from Ref.~\cite{bischoff_19}.}
\label{fig:cur_vol_struct}
\end{figure}

The current-voltage diagram is plotted in Fig.~\ref{fig:cur_vol_struct}(a).  The data show that in the low-voltage regime, the currents decrease as $\gamma / \omega_0$ is increased.  Similar behaviour is observed in the SAHM in Ref.~\cite{khedri_18}.  This is even more clearly illustrated in Fig.~\ref{fig:cur_vol_struct}(b), where $\expval*{\hat j(t)} _{\rm av}/V$ is plotted.  As $\gamma/\omega_0$ is increased, a steady-state current can not be estimated for large $V/\omega_0$ from our data. For this  reason, we show fewer points in those cases.  

In the SAHM, the equilibrium spectral function of the dot at the particle-hole symmetric point displays a main peak at zero frequency accompanied by additional peaks separated by $\omega_0$~\cite{jovchev_13}.  When the electron-phonon coupling is increased,  the width of the main peak starts decreasing as spectral weight is shifted to larger frequencies.  Since the current contains the integral over the spectral function (for the voltages studied here, one would expect the spectral function to remain approximately  unaffected), the current will decrease at small voltages.  Although the spectral function of the Holstein-structure extending over several sites is likely more complicated (see, e.g., Refs.~\cite{zhang99,sykora_05,zhao_05} for spectral functions of half-filled Holstein chains),  it is plausible that this picture still holds.  This is further supported by the fact that for the non-interacting model and at small voltages,  increasing $t_0/t_{\rm hyb} $ leads to a decrease in the current.  Decreasing $t_0/t_{\rm hyb} $, however,  increases the current.  This is because at a fixed small voltage, a larger portion of states of the structure participates in transport as the band width decreases. Note that if both   $t_0$ and $t_{\rm hyb} $ are decreased but their ratio is kept constant, the current also decreases. For the Holstein structure, one expects that both the tunneling in and out of the structure as well as the band width is reduced when electron-phonon interactions are turned on. Since we clearly see a decrease in the current, we assume that the dominating effect of the electron-phonon coupling is on the tunneling from the structure into the leads in the parameter regimes studied here.

In the inset of Fig.~\ref{fig:cur_vol_struct}(b), we show the Luttinger-liquid parameter computed from our data together with those obtained by studying the structure factor in Ref.~\cite{ejima09} and from the Kubo formalism in Ref.~\cite{bischoff_19}. The Luttinger parameter $K$  renormalizes  the conductance $G$ in a Luttinger liquid~\cite{apel_82,kane_92,kane_fischer_92},
\begin{equation} \label{def_condparam}
G=KG_0,
\end{equation}
where $G_0$ is the conductance of free fermions in a tight-binding chain. We calculate $K$ as the ratio of $\expval*{\hat j}_{\rm av}$ at finite $\gamma$ and at $\gamma=0$ with $V/ \omega_0=0.2$  \begin{equation} \label{def_K}
K=\frac{\expval*{\hat j}_{\rm av}}{\expval*{\hat j}_{\rm av, \gamma/ \omega_0=0}}\biggr\rvert_{V/ \omega_0=0.2}.
\end{equation}
Note that we changed the averaging interval to $t\omega_0 \in [25,32.5]([30,32.5])$ for $\gamma/\omega_0 =1.4 (1.5)$  due to the longer relaxation time.  We see that this method qualitatively reproduces $K$ from our  time-dependent calculations.

\begin{figure}[t]
\includegraphics[width=0.99\columnwidth]{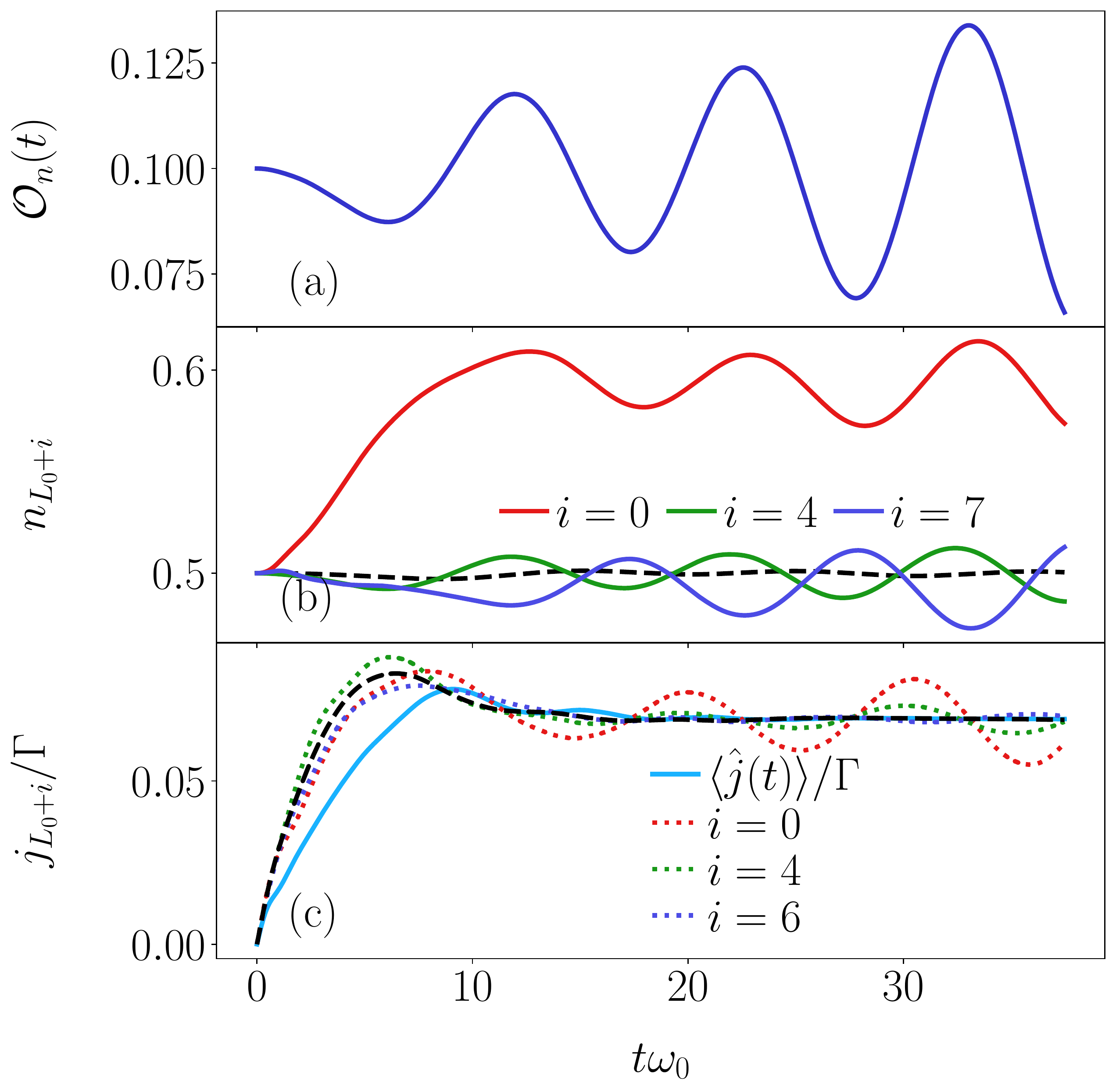}
\caption{(a) Order parameter in the electron sector,  see Eq.~\eqref{def_Opamn},  of the Holstein structure with $L_{\rm s}=9,L=236,    t_{l}/\omega_0=2,\gamma / \omega_0=1,  \Gamma /\omega_0=1,  \tilde{\epsilon}=0$ and $V / \omega_0=0.6$.  For the calculations, we use $\epsilon_{\rm LBO}=10^{-7},\epsilon_{\rm bond}=10^{-8}$.  (b) Selected local densities in the structure, see main text for details.  (c) Current from Eq.~\eqref{def_curr} together with selected local currents in the structure, see main text for details.  The black dashed lines are the average over local densities (b) and local currents (c).}
\label{fig:METdens}
\end{figure}

We now take a closer look at the current-carrying state at $V/ \omega_0=0.6$ and $\gamma / \omega_0=1$.  Figure~\ref{fig:METdens}(a) shows the order parameter, which oscillates around $0.1$.  Further,  the local electron densities $n_{L_0+i}=\expval*{\hat n_{L_0+i}}$ on selected sites are plotted in Fig.~\ref{fig:METdens}(b).  As expected, their average increases for $i$ going from small to large due to the inhomogeneous bias voltage. Still,  their mean value is $1/L_{\rm s} \sum\limits_{i=0}^{L_{\rm s }-1}\expval*{\hat n_{L_0+i}}\approx 0.5$, as illustrated by the black dashed line.  Selected local currents $j_{L_0+i}=i  \expval*{\hat c_{L_0+i}^{\dagger}\hat c_{L_0+i+1}-h.c. }$ and the total current, see Eq.~\eqref{def_curr}, can be seen in Fig.~\ref{fig:METdens}(c).  There, the steady-state current and the mean of the local currents (black dashed line) overlap after some initial dynamics, consistent with having a constant flow of current through the structure. The local currents  all oscillate around the mean value.  Both the local currents and densities are also representative for those not shown in this paper.
\subsubsection{Reduced density matrices in the metallic phase  }
\label{subsubsec:metreddm1}

\begin{figure}[t]
\includegraphics[width=0.99\columnwidth]{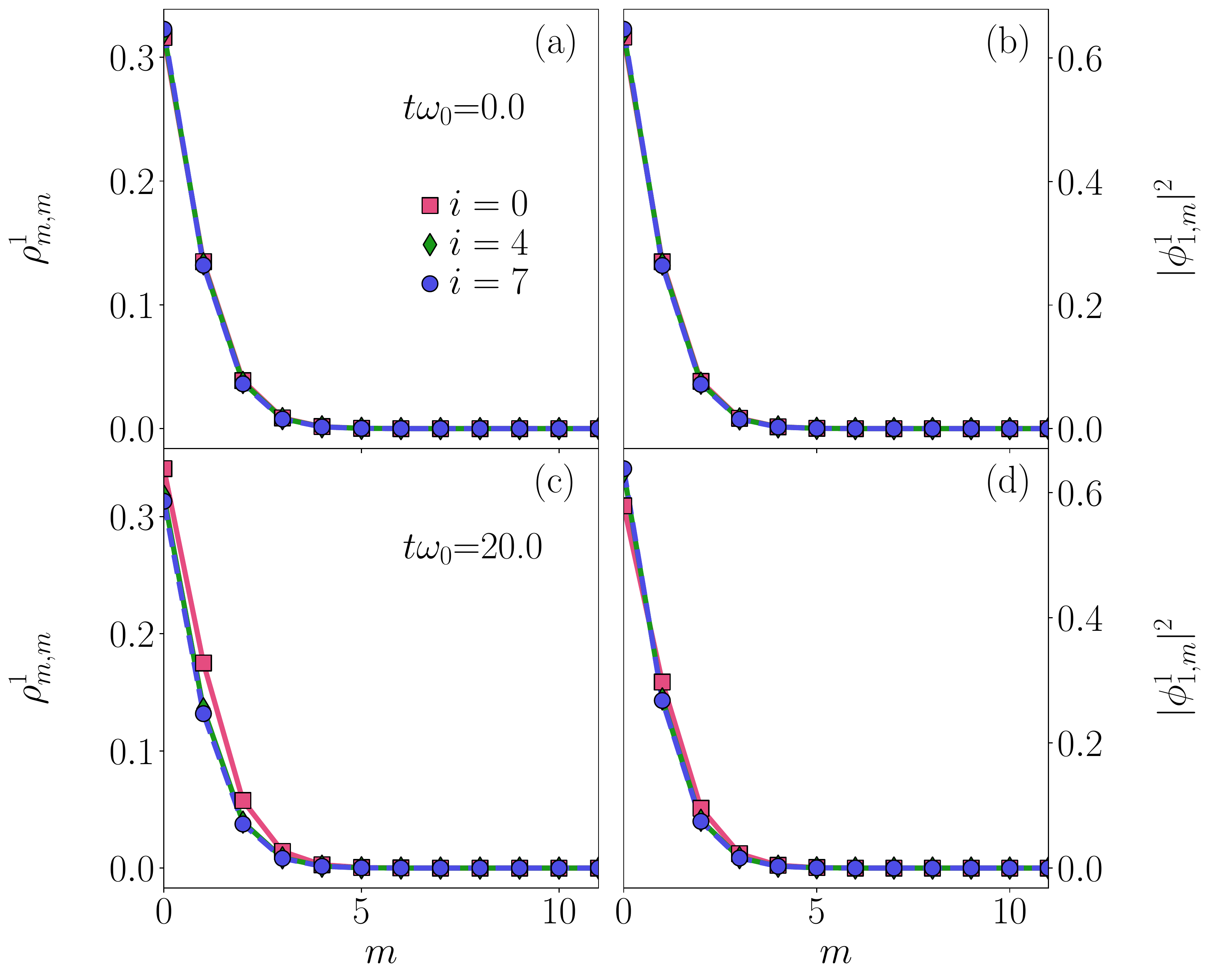}
\caption{(a) Diagonal elements of the reduced density matrix in  the Holstein structure with $L_{\rm s}=9,L=236,    t_{l}/\omega_0=2,\gamma / \omega_0=1,  \Gamma /\omega_0=1,  \tilde{\epsilon}=0,M=30$ at different sites and different times with $ V / \omega_0=0.6$.  Further, $\epsilon_{\rm LBO}=10^{-7}$ and $\epsilon_{\rm bond}=10^{-8}$ are used for the calculations.  (b)  Absolute value squared of the components of the most significant optimal-basis state.  All data are from the block matrix with an electron and the times are $t\omega_0=0$ in (a) and (b), and $t\omega_0=20$ in (c) and (d).}
\label{fig:localmet}
\end{figure}

Lastly, we look at the diagonal elements of the reduced density matrix in the one-electron sector $\rho^1$ and the most significant eigenvector $\rho^1 \ket*{\phi^1_1}=w^1_1 \ket*{\phi^1_1}$ at site $L_0+i$, where $w_1^1$ is the largest eigenvalue.  The diagonal elements of $\rho$ were already studied for the SAHM in different parameter regimes in Ref.~\cite{eidelstein_13} and the optimal-basis states in the Holstein model in, e.g., Ref.~\cite{dorfner_vidmar_15}. Both are shown for selected sites for $V/ \omega_0=0.6$ and $\gamma / \omega_0=1$ at different times in Fig.~\ref{fig:localmet}. We see that both the diagonal elements of the reduced density matrix and the component of the optimal basis are strongly peaked at the zero phonon mode and decay rapidly for larger modes.  We also observe that the distributions remain approximately the same during the time evolution (here, we only show the data for $t \omega_0=0,20$) with some oscillations for the different sites, stemming from the oscillating local densities.  Physically, this means that the electrons are being transported through the structure without significantly impacting the phonon distributions.  Indeed, our results indicate that the current-carrying state can be well described by a few local modes that do not display much  change compared to the ground state. The most significant eigenvalues of the reduced density matrix can be seen  in Appendix~\ref{sec:app1} and decay exponentially. This illustrates the computational benefit of  using LBO. 

\subsection{Charge-density wave phase}
\label{subsec:cdwpahse}
\subsubsection{Charge-density wave breakdown  }
\label{subsubsec:cdwbreak}
  \begin{figure}[t]
\includegraphics[width=0.99\columnwidth]{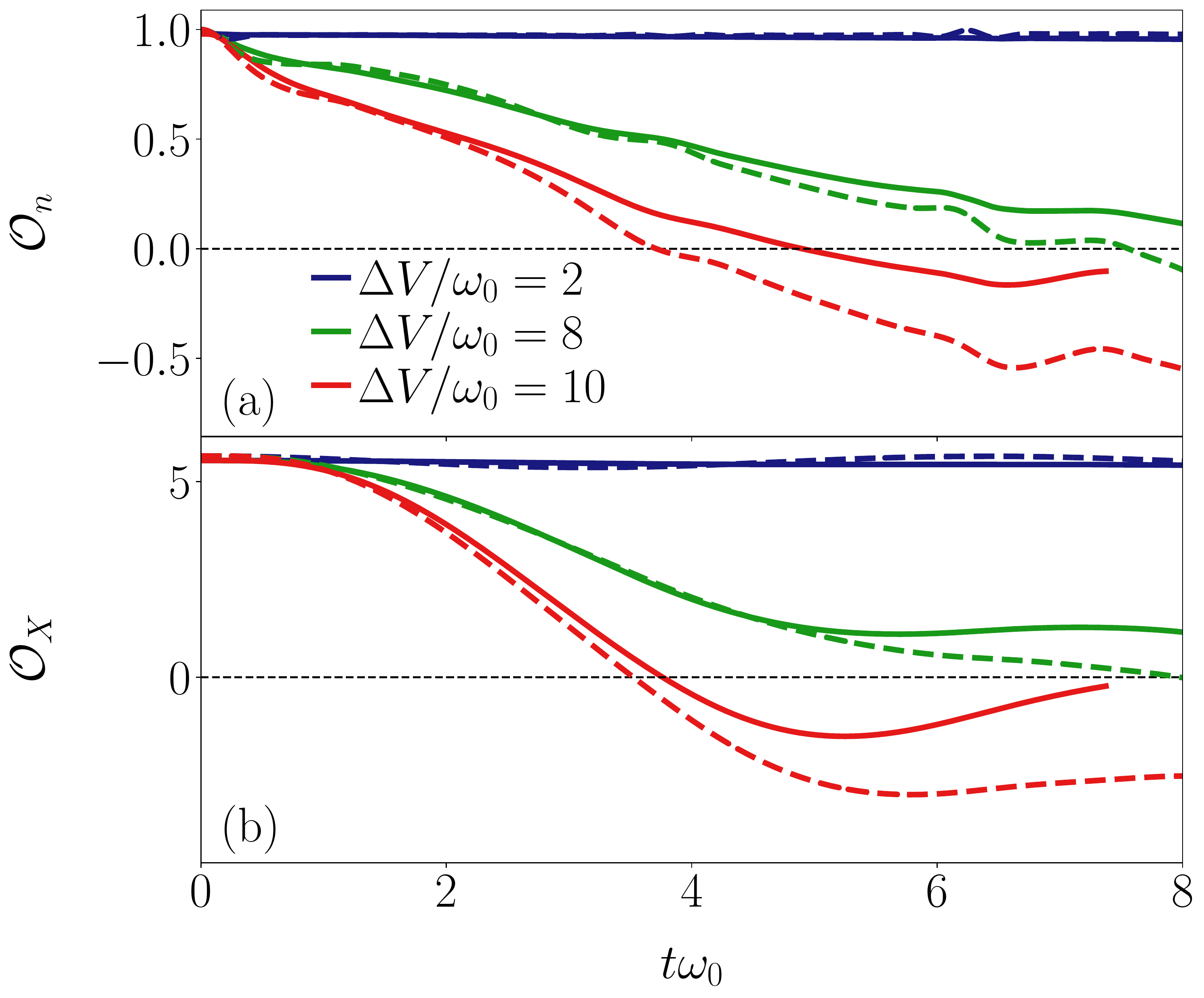}
\caption{Order parameters for the Holstein structure with $L_{\rm s}=9,L=236,    t_{l}/\omega_0=2,\gamma / \omega_0=2\sqrt{2},  \Gamma /\omega_0=1,  \tilde{\epsilon}=0$ and voltage gradients with different $\Delta V / \omega_0$.   For the calculations we use $\epsilon_{\rm LBO}=10^{-7}$ and $\epsilon_{\rm bond}=10^{-7}$. (a) Order parameter in the fermion sector, see Eq.~\eqref{def_Opamn}.  (b) Order parameter in the phonon sector, see Eq.~\eqref{def_OpamX}.  The dashed lines show the exact data for the Holstein dimer. }
\label{fig:structOP}
\end{figure}
 
 \begin{figure}[t]
\includegraphics[width=0.99\columnwidth]{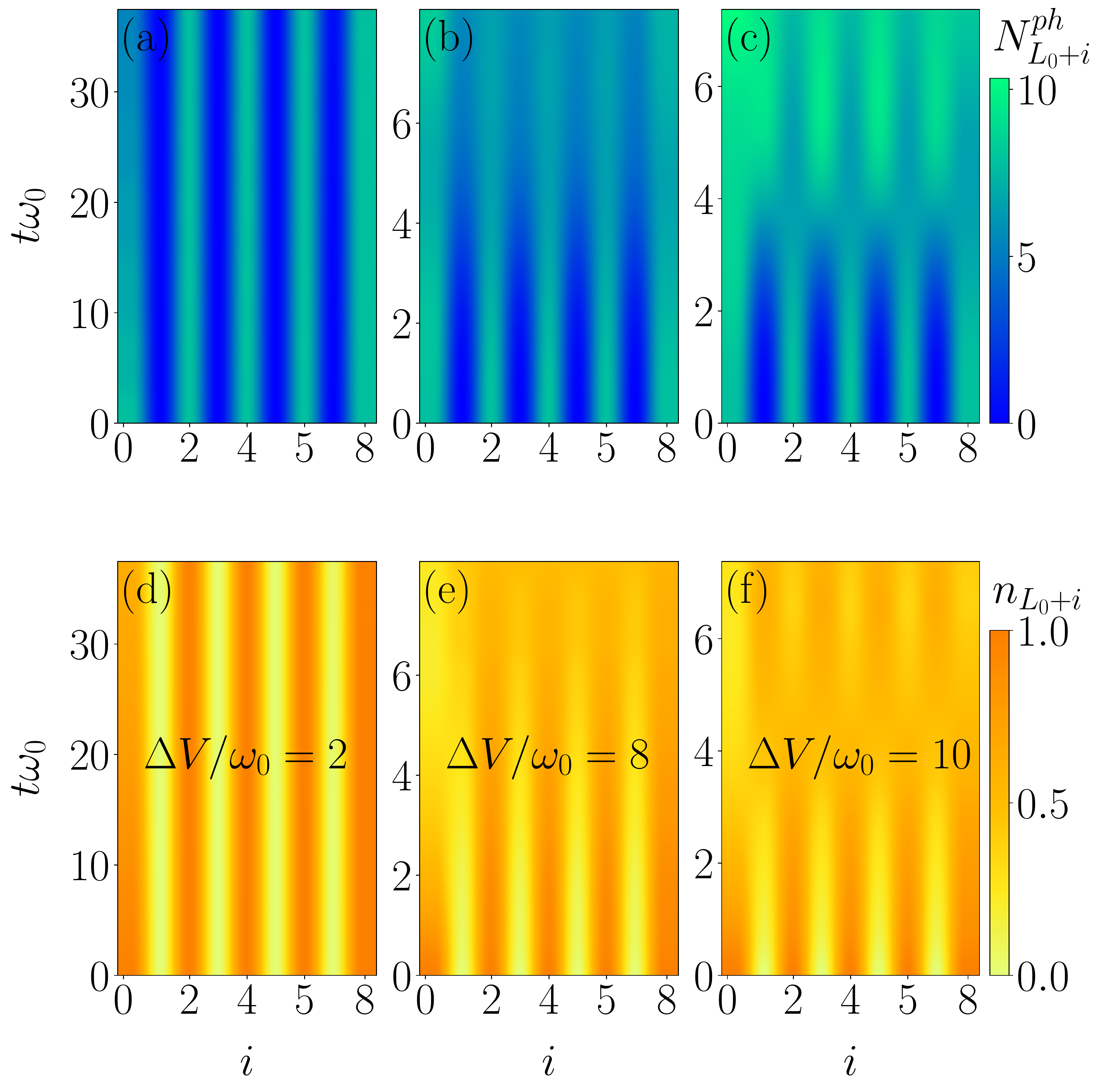}
\caption{Local densities for the Holstein structure with $L_{\rm s}=9,L=236,    t_{l}/\omega_0=2,\gamma / \omega_0=2\sqrt{2},  \Gamma /\omega_0=1,  \tilde{\epsilon}=0,M=50$ and different local voltage differences $\Delta V / \omega_0$.   We use $\epsilon_{\rm LBO}=10^{-7}$ and $\epsilon_{\rm bond}=10^{-7}$ for the calculations.  (a)-(c) Local phonon number for $\Delta V/ \omega_0=2,8,10$.  (d)-(f) Local electron occupation for the same  values of $\Delta V$.}
\label{fig:CDWdens}
\end{figure}

We now go into the strong electron-phonon coupling (CDW) regime where the order parameters, see Eqs.~\eqref{def_Opamn} and~\eqref{def_OpamX},  remain nonzero as $L_{\rm s}$ is increased.  Since the polaron binding energy is $\gamma^2/\omega_0$, we expect that voltages with $\Delta V\sim \gamma^2/\omega_0$ are needed to break up the CDW as we saw in Sec.~\ref{sec:holmod}.  Here, we also choose $\gamma/\omega_0=2\sqrt{2}$ and study the breakdown of the CDW at large $V/\omega_0$.  In Fig.~\ref{fig:structOP}, we show the order parameters for $\Delta V/ \omega_0 =2,8,10$.  We observe a similar behaviour as for the regular Holstein model, namely that they decrease as the applied voltage is increased. Further,  we see that the initial dynamics are quantitatively similar to those of  the dimer.  However,  as was the case in Sec.~\ref{sec:holmod}, their decay is qualitatively different, in particular for larger times.  Whereas the dimer data go far below zero,  our data indicate that for the structure, both order parameters decay to zero with a stronger damping. While not shown here, when fitting the decay of the order parameters with a linear function for the initial dynamics, as done in Sec.~\ref{sec:holmod}, we see a similar behaviour as for the Holstein model. The difference  is that no rescaling is needed since all electrons can contribute to the decay.  Further, the data points for the structure tend to lie above the dimer points due to boundary effects which we elaborate on in the next paragraph.

The local electron and phonon occupations are shown as functions of the sites in the structure and of time  in Fig.~\ref{fig:CDWdens}.  The figure confirms the results from the order parameters by illustrating how the electron densities remain constant for $\Delta V/\omega_0=2$ [Fig.~\ref{fig:CDWdens}(d)] and start to spread out for $\Delta V/\omega_0=8,10$ [Fig.~\ref{fig:CDWdens}(e) and (f)].  Further,  the change in electron densities is accompanied by a change and a shift of the maximum phonon occupation to the previously empty sites.  Indeed, once the electrons can tunnel to an empty site, the excess energy first goes into generating a large number of new phonons. Additionally, the coupling to the leads allow the electron furthest to the left to tunnel out of the structure. This boundary effect also leads to a small decay of the order parameter for $\Delta V/ \omega_0=2$ as well. We also observe that there are no local currents present for $\Delta V /\omega_0=2 $, but that they become finite for large $\Delta V /\omega_0$.
\subsubsection{Reduced density matrices in the charge-density wave phase }
\label{subsubsec:cdwreddm}
\begin{figure}[t]
\includegraphics[width=0.99\columnwidth]{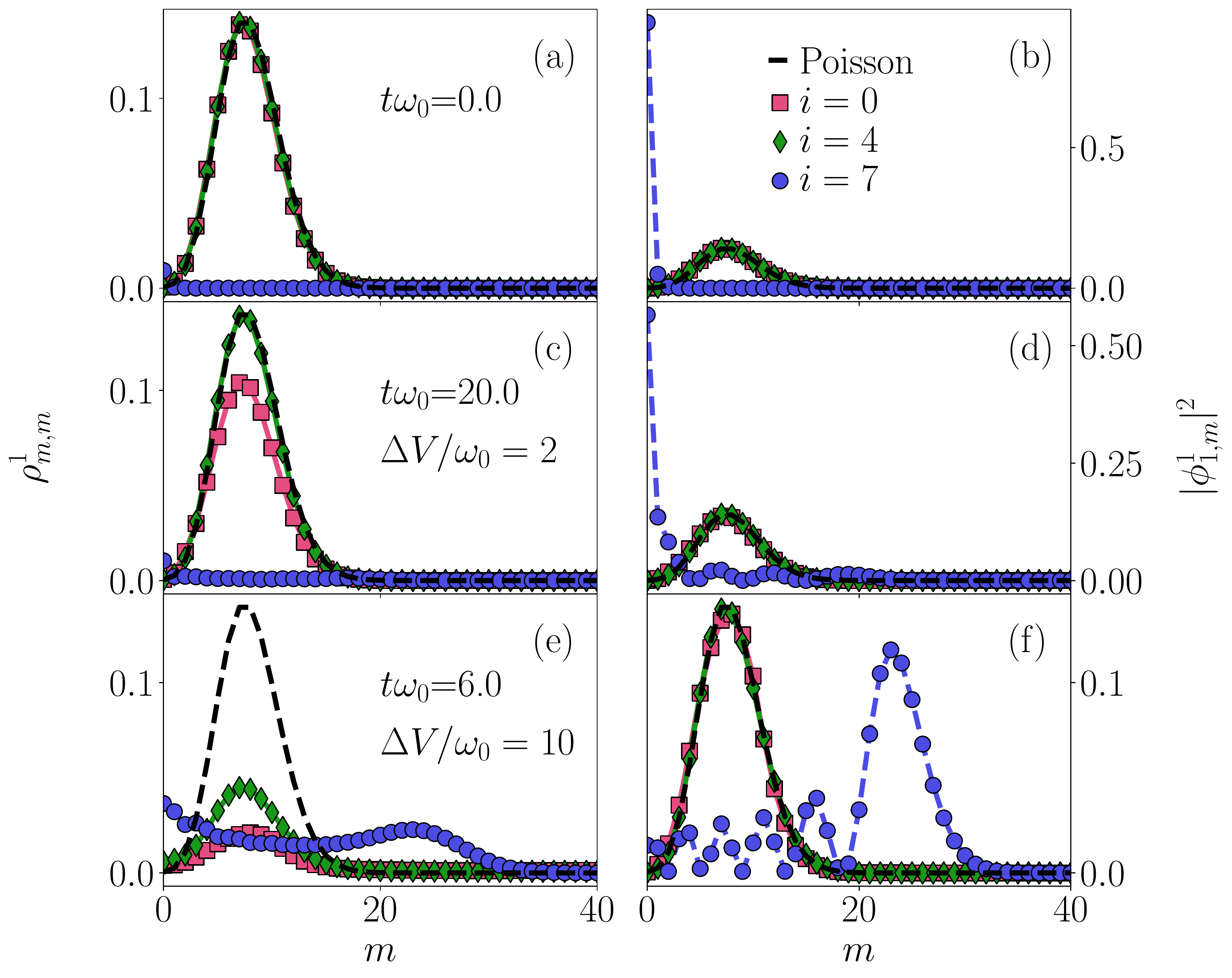}
\caption{(a) Diagonal elements of the reduced density matrix in the Holstein structure with $L_{\rm s}=9,L=236,    t_{l}/\omega_0=2,\gamma / \omega_0=2\sqrt{2},  \Gamma /\omega_0=1,  \tilde{\epsilon}=0,M=50$.  (b) Components of the most significant local optimal-basis state. (c) Same as in (a) but at $t\omega_0=20$ and $\Delta V / \omega_0=2$.   (d) Same as in (a) but at $t\omega_0=6$ and $\Delta V / \omega_0=10$.  (d) Same as in (b) but at $t\omega_0=20$ and $\Delta V / \omega_0=2$.   (f) Same as in (b) but at $t\omega_0=6$ and $\Delta V / \omega_0=10$.   We use  $\epsilon_{\rm LBO}=10^{-7}$ and $\epsilon_{\rm bond}=10^{-7}$ for the calculations.  The black dashed line corresponds to the Poisson distribution from Eq.~\eqref{eq:posidis}.  }
\label{fig:cdwlbo1}
\end{figure}

We continue by looking at the diagonal elements of the reduced density matrix and the first optimal-basis state and contrast them to what we observed for the current carrying state in Sec.~\ref{subsec:metal}.  In Fig.~\ref{fig:cdwlbo1}, we show the data for different voltages at different times. The first thing that stands out are the initial distributions at $t\omega_0=0$ in Figs.~\ref{fig:cdwlbo1}(a) and (b). On the sites occupied by electrons, namely sites $i=4,7$, both the diagonal elements of the density matrix and the most significant optimal-basis state are well described by the Poisson distribution, indicating a coherent local phonon state, as expected in the large-coupling limit. To illustrate this,  we also plot
     \begin{equation} \label{eq:posidis}
P_{\rm Poisson}(m)=\frac{\abs{\lambda }^{2m}}{m!} e^{-\abs{\lambda }^2}\, ,
\end{equation} where $\lambda=\gamma / \omega_0$.
The empty sites have almost no weight except for at the $m=0$ mode.  As the system evolves in time with a small bias voltage we make several interesting observations. As illustrated in Figs.~\ref{fig:cdwlbo1}(c) and (d), at $t\omega_0=20$, the system remains well described by coherent states. Both the most significant optimal-basis state and the diagonal elements of the density matrix keep their Poisson form. Whereas $\abs*{\phi^1_{1,m}}^2$ remains perfectly described by the Poisson distribution on the occupied sites, the amplitude of the $\rho^1_{m,m}$ at $i=0$ decreases due to the boundary effect previously described.  We further note some change in $\abs*{\phi^1_{1,m}}^2$ for $i=4$.

More dramatic changes can be seen for the large bias voltage $\Delta V/ \omega_0=10$ in Fig.~\ref{fig:cdwlbo1}(e) and (f).  At $t\omega_0=6$, the amplitude of the diagonal elements of $\rho^1$ have decreased significantly compared to the original distribution at sites $i=0,4$. This is a consequence of the electron density getting distributed to other sites, leading to an increase of the weights in the zero-electron density matrix $\rho^0$, which is not shown here.  Also, $\abs*{\phi^1_{1,m}}^2$ remains well described by the coherent state on all initially occupied sites, but with some oscillations between modes. However, the previously empty sites gain a large amplitude spreading out across several modes as can be seen for $i=7$.  This is due to the electron density increasing and phonons being generated with the excess energy. Further, the $\abs*{\phi^1_{1,m}}^2$ seems to resemble a shifted coherent state with additional oscillations at the times reachable here.  Figures~\ref{fig:cdwlbo1}(e) and (f) also illustrate that the system is notoriously more complex to simulate due to the large number of bare modes needed to capture the dynamics on the initially unoccupied sites.  We further looked at the second most significant optimal-basis state in the one-electron sector, but there, no physical interpretation could be extracted. In Appendix~\ref{sec:app1}, we additionally illustrate that a large number of optimal modes are needed to keep the error small and thus LBO loses some of its advantage. 

\section{Conclusion} \label{sec:conclusion}
In this paper, we used time-dependent DMRG with local basis optimization to investigate both the Holstein model in the CDW phase and the Holstein model sandwiched between conducting leads in the CDW and TLL phase under an applied bias voltage.  For the regular Holstein model, we first saw that when a strong enough bias voltage is applied,  the order parameters in both the electron and phonon sector clearly decay. We further demonstrated a clear dependence between the applied voltage and the initial decay rate of the order parameters and that the dynamics resemble those in the Holstein dimer for short times but later deviate.

We then proceeded to look at the Holstein model coupled to conducting leads. After  establishing  that a clear separation between the TLL and CDW phase can be seen at different coupling strengths we first focused  on the system in the TLL phase. Studying the steady-state currents for small voltages, we were able to compute  current-voltage curves, showing that charge transport at small voltages is reduced when the phonon coupling is increased. Similar behaviour has been reported for the SAHM model by Ref.~\cite{khedri_18}.  Our data also reproduce the Luttinger-liquid  parameters from Refs.~\cite{ejima09,bischoff_19}. We additionally looked at the diagonal elements and the most significant optimal-basis state in the one-electron sector for the current-carrying state.  Our data indicate that the local phonon distribution is not significantly impacted by the applied voltage in  the steady-state current. Further,  the same is seen for the most significant optimal mode.  

We then continued our study of the Holstein model coupled to leads by working with an  electron-phonon coupling in the CDW regime. We could report similar behaviour as for the regular Holstein model, namely that large voltages are needed for the CDW to break down. At small voltages, the CDW remains unchanged except for boundary effects. This is also seen in the properties of the reduced density matrices.
Initially,  the occupied sites were well described by the Poisson distribution. This remained true during the time evolution for small bias voltages, except for the aforementioned boundary effects. However, for strong voltages, the amplitudes of the reduced density matrix in the one-electron sector, $\rho^1$, started to decrease as expected since the electron densities spread out. Further, a wide range of bare modes displayed large weights in the previously unoccupied sites as a consequence of the phonons being generated with the excess energy.
In this case, the most significant optimal-basis state remain well described by the coherent distribution for all times reached here in the initially occupied sites. In contrast, the previously empty sites obtained a distribution centered around the larger bare phonon modes.

There are many possible continuations building on the results of this work. After having established that local basis optimization can simplify the calculations significantly in the metallic phase,  one could make the structure more complex by either adding phonon dispersion (see, e.g., Ref.~\cite{costa_18,bonca_21}) or trying to capture more aspects  of manganites (see, e.g., Ref.~\cite{ifland_15,kalla_19} and Ref.~\cite{hotta_04} for a theory review) by adding electron interaction or interaction with local spins.  Also, analysing how energy is transferred into the different degrees of freedom, complementary to Ref.~\cite{weber_2021}, would be of great interest. Further, applying a thermal gradient to the system would allow for the study of thermal transport through such quantum structures, see, e.g., Refs.~\cite{koch_14,kherdi_meden_17,khedri_18,schinabeck_18,Roura_Bas_19,shi_20}. Lastly, being able to compute a current-voltage diagram for the CDW phase, its dynamics, and its properties at finite temperature would be of great interest, e.g., in the context of recent experiments~\cite{wahl_03,Jooss_07,barone_09}.  In particular, it would be desirable to see if one can reach a state where the electron densities are equally distributed through the system and to better understand the time scales of any collective behaviour.  Since the reachable times of our CDW calculations are greatly limited by the large number of optimal-basis states required,  this might be a task for other phonon-specialized matrix-product state based time-evolution techniques~\cite{mardazad_21}. Here, also insights into different behaviours of CDW and CO  states are of interest.
\section*{Acknowledgement} \label{sec:ack}
We  acknowledge  useful  discussions with P. Blöchl,  J.  Hoffmann,  E. Jeckelmann,  B. Kressdorf,  V. Meden, and J. Stolpp.
This work was funded by the Deutsche Forschungsgemeinschaft (DFG, German Research Foundation) – 217133147 via SFB 1073 (project B09 and project B02).  We further thank E. Jeckelmann and V.  Meden for providing us data to benchmark our results from Refs.~\cite{bischoff_19} and~\cite{khedri_18}, respectively. 

\appendix
\section{Numerical details} \label{sec:app1}
\begin{figure}[t]
\includegraphics[width=0.99\columnwidth]{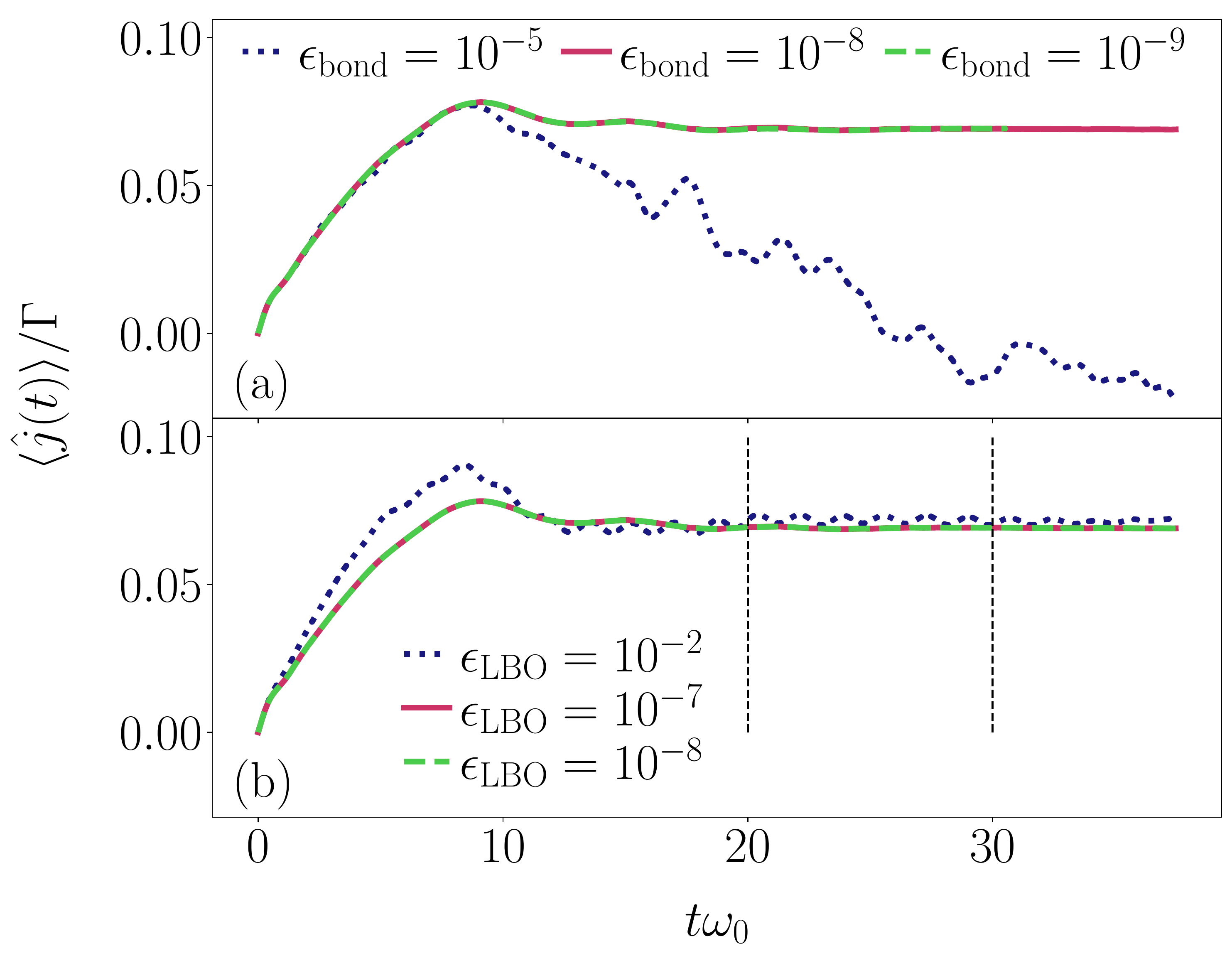}
\caption{Current $\expval*{\hat j(t)}$ from Eq.~\eqref{def_curr} for the Holstein structure with $L_{\rm s}=9,L=236,  V/ \omega_0=0.6,  t_{l}/\omega_0=2,\Gamma /\omega_0=1,  \tilde{\epsilon}=0,M=30$ and $\gamma / \omega_0=1$.  (a) Fixed $\epsilon_{\rm LBO}=10^{-7}$ and different $\epsilon_{\rm bond}$. (b) Fixed $\epsilon_{\rm bond}=10^{-8}$ and different values of $\epsilon_{\rm LBO}$.}
\label{fig:steadystate_struct}
\end{figure}

Here, we illustrate several numerical details of our calculations. As explained in Sec.~\ref{sec:meth}, we control the truncation of the bond dimension with the parameter $\epsilon_{\rm bond}$ and the truncation of the local basis optimization with $\epsilon_{\rm LBO}$.  Figure~\ref{fig:steadystate_struct} shows the expectation value of the current for fixed $\epsilon_{\rm bond}$ and different $\epsilon_{\rm LBO}$ in (a) and fixed $\epsilon_{\rm bond}$ and different $\epsilon_{\rm LBO}$ in (b). Clearly, the current is converged for the parameters used in this work, which also was verified for the other data shown. Additionally, we observe that using a very large $\epsilon_{\rm bond}$ unequivocally leads to the false expectation values for $\expval*{\hat j(t)}$.  In contrast, very large $\epsilon_{\rm LBO} $ can be used.  Further,  even a  too large $\epsilon_{\rm LBO} $ seems to approximately reproduce the steady-state current for the parameters shown here, and therefore,  more care must be taken when determining if $\epsilon_{\rm LBO} $ is sufficiently converged.  To compute the current-voltage diagram we average $\expval*{\hat j(t)}$ in the interval $t\omega_0 \in [20,30]$, which is illustrated by the black dashed lines in Fig~\ref{fig:steadystate_struct}(b).

\begin{figure}[t]
\includegraphics[width=0.99\columnwidth]{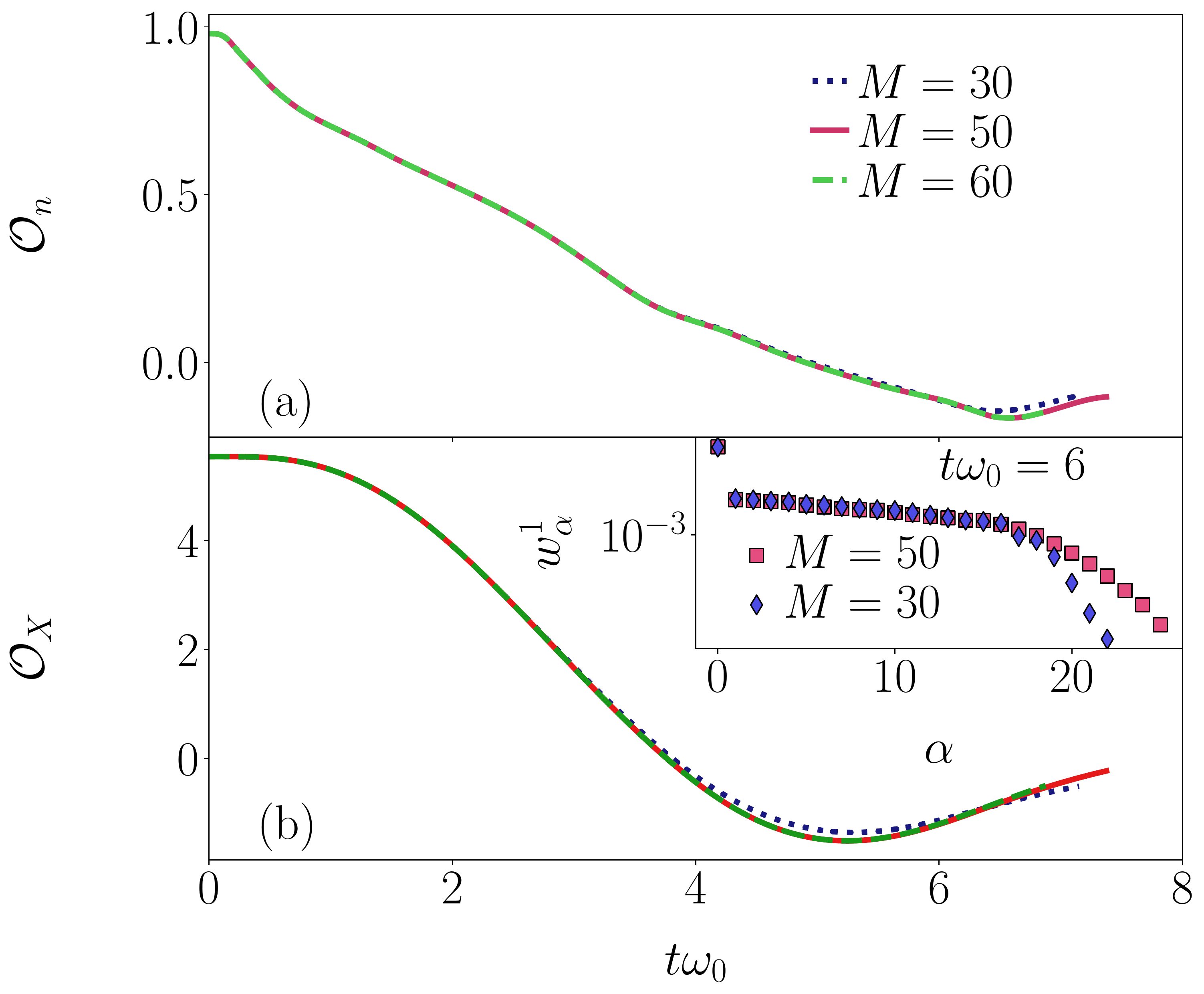}
\caption{Order parameters  for the Holstein structure with $L_{\rm s}=9,L=236,    t_{l}/\omega_0=2,\gamma / \omega_0=\sqrt{2},  \Gamma /\omega_0=1,  \tilde{\epsilon}=0,\Delta V / \omega_0=10$, and different $M$.   For the calculations, we use $\epsilon_{\rm LBO}=\epsilon_{\rm bond}=10^{-7}$. (a) Order parameter in the fermion sector, see Eq.~\eqref{def_Opamn}.  (b) Order parameter in the bosonic sector, see Eq.~\eqref{def_OpamX}.  The inset shows the eigenvalues of the reduced density matrix in the one-electron sector at time $t\omega_0=6$ on a log-scale.}
\label{fig:structOPdM}
\end{figure}

To demonstrate that sufficiently large local phonon Hilbert spaces are included in the calculations we show the order parameter computed with different local phonon-number truncations $M$ in Fig.~\ref{fig:structOPdM}. For the parameters shown here, $M=50$ is more than sufficient to capture the relevant physics of the order-parameter decay.  Further, we see that the $M=30$ data start to deviate on the scale of the figure.  Also, inspecting the optimal weights in the inset of Fig.~\ref{fig:structOPdM} reveals that $M=30$ and $M=50$ differ with respect to the larger optimal-basis state weights $w^1_{\alpha}$ in the one-electron sector.

\begin{figure}[t]
\includegraphics[width=0.99\columnwidth]{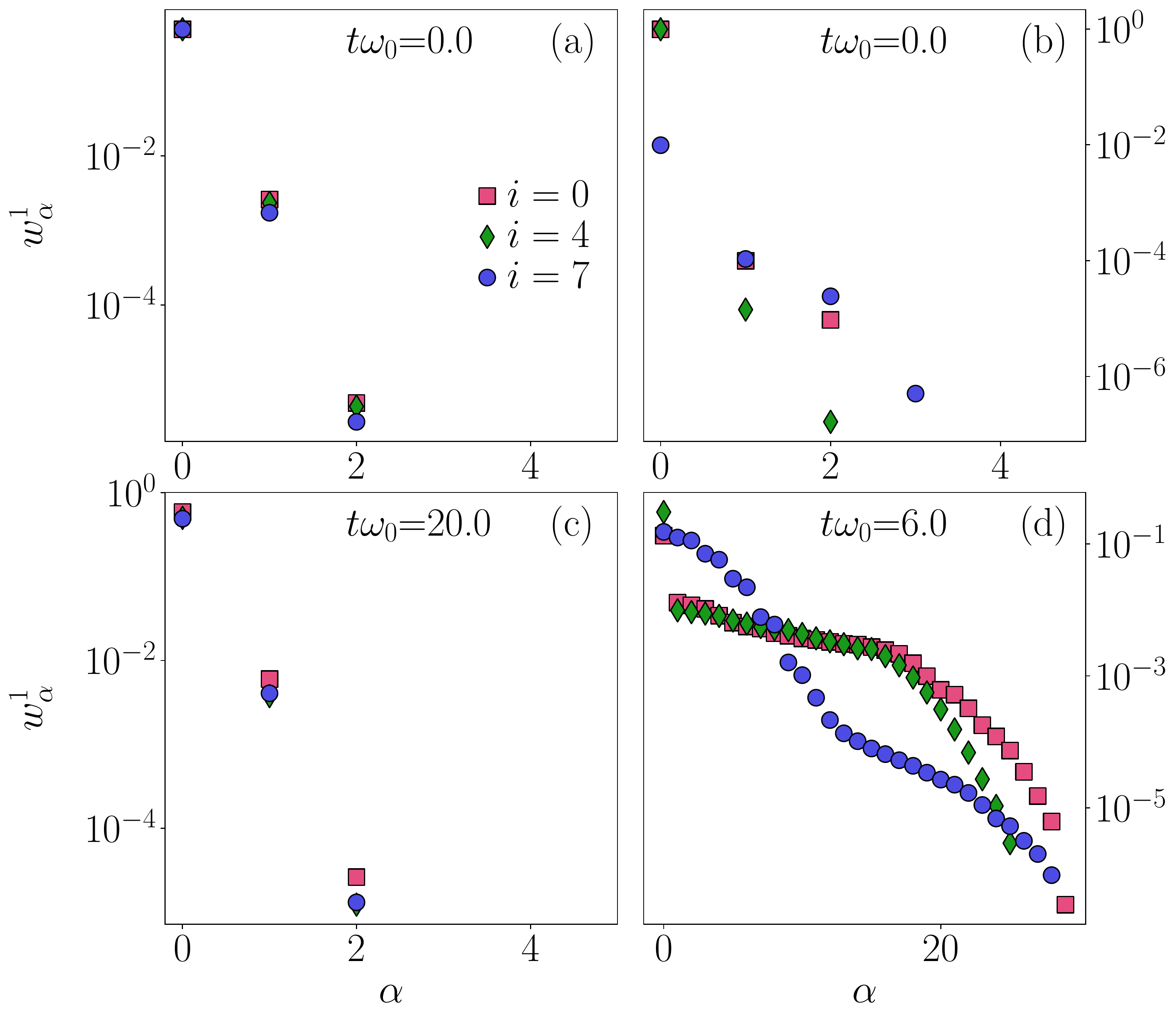}
\caption{Eigenvalues of the reduced density matrices at different sites $i$ for the Holstein structure with $L_{\rm s}=9,L=236,   t_{l}/\omega_0=2,\Gamma /\omega_0=1,  \tilde{\epsilon}=0,$ and different $\gamma / \omega_0$  and $M$ and at different times. (a) $t\omega_0=0,M=30$ and $\gamma / \omega_0=1$.  (b) $t\omega_0=0,M=50$ and $\gamma / \omega_0=2\sqrt{2}$. (c) Same as (a) but at $t\omega_0=20$ and $V/\omega_0=0.6$.  (d) Same as (b) but at $t\omega_0=6$ and $V/\omega_0=100$ $(\Delta V/\omega_0=10)$.  We use  $\epsilon_{\rm LBO}=10^{-7}$ in all plots and $\epsilon_{\rm bond}=10^{-8}[10^{-7}]$ in (a) and (c)  [(b) and (d)].  We only show $w^1_{\alpha}>\epsilon_{\rm LBO}$.}
\label{fig:weights}
\end{figure}

In Fig.~\ref{fig:weights}, we show the eigenvalues $w^1_{\alpha}$ of the reduced density matrix $\rho^1$ for different sites $i$ on a logarithmic scale.   In the TLL regime, displayed in Figs.~\ref{fig:weights}(a) and (c), it becomes clear that LBO can be a powerful tool to further study the current-carrying state.  For all times calculated here, only a few optimal modes are needed to accurately represent it.  The situation for the CDW breakdown is quite different. Figures \ref{fig:weights}(b) and (d) show the weights in this regime. Despite being able to represent the state accurately with only a few modes at $t\omega_0=0$, almost the complete set of modes is needed for $t\omega_0=6$.  This could in theory make  the optimal-basis calculations even more costly than just the regular time-evolution method.  However, that is not observed for our calculations with $M=50$. This regime is still clearly a candidate for other schemes to efficiently treat phonons, such as the one of Ref.~\cite{koehler20}.  In total,  the limiting factor for our calculations in the metallic regime is the local bond dimension and obtaining a steady state.  For the CDW regime, the amount of local modes needed becomes an issue before the bond dimension matters. 

\bibliographystyle{biblev1}
\bibliography{references}

\end{document}

%% file: sketch.pdf_tex
\begingroup%
  \makeatletter%
  \providecommand\color[2][]{%
    \errmessage{(Inkscape) Color is used for the text in Inkscape, but the package 'color.sty' is not loaded}%
    \renewcommand\color[2][]{}%
  }%
  \providecommand\transparent[1]{%
    \errmessage{(Inkscape) Transparency is used (non-zero) for the text in Inkscape, but the package 'transparent.sty' is not loaded}%
    \renewcommand\transparent[1]{}%
  }%
  \providecommand\rotatebox[2]{#2}%
  \newcommand*\fsize{\dimexpr\f@size pt\relax}%
  \newcommand*\lineheight[1]{\fontsize{\fsize}{#1\fsize}\selectfont}%
  \ifx\svgwidth\undefined%
    \setlength{\unitlength}{20.06255077bp}%
    \ifx\svgscale\undefined%
      \relax%
    \else%
      \setlength{\unitlength}{\unitlength * \real{\svgscale}}%
    \fi%
  \else%
    \setlength{\unitlength}{\svgwidth}%
  \fi%
  \global\let\svgwidth\undefined%
  \global\let\svgscale\undefined%
  \makeatother%
  \begin{picture}(1,0.75716698)%
    \lineheight{1}%
    \setlength\tabcolsep{0pt}%
    \put(64.51856455,-43.48660483){\color[rgb]{0,0,0}\makebox(0,0)[lt]{\begin{minipage}{25.54774603\unitlength}\raggedright \end{minipage}}}%
    \put(0,0){\includegraphics[width=\unitlength,page=1]{sketch.pdf}}%
    \put(0.22103408,0.59900902){\color[rgb]{0,0,0}\makebox(0,0)[lt]{\lineheight{1.25}\smash{\begin{tabular}[t]{l}$\omega_0$\end{tabular}}}}%
    \put(0.07853234,0.40332045){\color[rgb]{0,0,0}\makebox(0,0)[lt]{\lineheight{1.25}\smash{\begin{tabular}[t]{l}$t_{\rm l}$\end{tabular}}}}%
    \put(0.48058446,0.17902116){\color[rgb]{0,0,0}\makebox(0,0)[lt]{\lineheight{1.25}\smash{\begin{tabular}[t]{l}$t_0$\end{tabular}}}}%
    \put(0.70495364,0.40332045){\color[rgb]{0,0,0}\makebox(0,0)[lt]{\lineheight{1.25}\smash{\begin{tabular}[t]{l}$t_{\rm hyb}$\end{tabular}}}}%
    \put(0.61890784,0.21003678){\color[rgb]{0,0,0}\makebox(0,0)[lt]{\lineheight{1.25}\smash{\begin{tabular}[t]{l}$\epsilon_b$\end{tabular}}}}%
    \put(0.33095804,0.37096475){\color[rgb]{0,0,0}\makebox(0,0)[lt]{\lineheight{1.25}\smash{\begin{tabular}[t]{l}$\gamma$\end{tabular}}}}%
    \put(0.1392269,0.02946574){\color[rgb]{0,0,0}\makebox(0,0)[lt]{\lineheight{1.25}\smash{\begin{tabular}[t]{l}$-V/2$\end{tabular}}}}%
    \put(0.8445486,0.68406511){\color[rgb]{0,0,0}\makebox(0,0)[lt]{\lineheight{1.25}\smash{\begin{tabular}[t]{l}$+V/2$\end{tabular}}}}%
  \end{picture}%
\endgroup%